\documentclass[a4paper,11pt]{article}
\pdfoutput=1 

\usepackage{jcappub} 

\usepackage[T1]{fontenc} 

\usepackage{amsmath}
\usepackage{hyperref}

\usepackage[english]{babel}
\usepackage{graphicx}
\usepackage{latexsym}
\usepackage{amsmath,amssymb}
\usepackage{natbib}
\usepackage{dcolumn}
\usepackage{bm}
\usepackage{calrsfs}
\usepackage{ulem}
\usepackage[usenames]{color}
\usepackage{xspace}

\newcommand{\be}{\begin{equation}}
\newcommand{\ee}{\end{equation}}
\newcommand{\nn}{\mbox{} \nonumber \\ \mbox{} }
\newcommand{\ba}{\begin{eqnarray}}
\newcommand{\ea}{\end{eqnarray}}
\newcommand{\om}{\omega}

\renewcommand{\v}{{\bf v}}

\newcommand\eg{{\it{e.g.}}}

\newcommand{\Bf}{{magnetic field}}
\newcommand{\Bfs}{{magnetic fields}}

\newcommand{\NS}{neutron star}
\newcommand{\NSs}{{neutron stars}}

\newcommand{\ms}{magnetosphere}
\newcommand{\mss}{magnetospheres}

\newcommand{\Lf}{Lorentz factor}
\newcommand{\WD}{White Dwarf}

\title{\boldmath Relativistic van Allen belts in magnetospheres of  pulsars and white dwarfs}

\author[a]{Maxim V.~Barkov\note{Corresponding author.}}
\author[b]{and Maxim Lyutikov}

\affiliation[a]{Institute of Astronomy, Russian Academy of Sciences,\\ Moscow, 119017 Russia}
\affiliation[b]{Department of Physics and Astronomy, Purdue University,\\ 525 Northwestern Avenue, West Lafayette, IN 47907-2036, USA}

\emailAdd{barkov@inasan.ru}

\abstract{We consider dynamics and  multi-frequency emission patterns  of relativistic van Allen belts - 
 particles trapped in the \mss\ of \NSs\  and {\WD}s. We account for   synchrotron  radiative losses and effects of relativistic  beaming of radiation. 
The system is non-Hamiltonian (non-energy conserving):  this results in  a wide non-scalable variety of spectral and temporal behaviors. There are  three  types of trapped particles' trajectories: (i) oscillating (particles experience multiple  bounces between magnetic bottles); (ii)  precipitating  (particles fall onto the star with finite transverse momentum); (iii) freezing (particles lose their transverse motion before  falling onto the star). 
The separation between  regimes (i) and (ii)  depends both on the ratio of the  bounce time to 
cooling time at magnetic equator $\tau_{ 0} $,    $\eta _0 =  R_0/( c \tau_0) \leq 1$,  as well as the  initial pitch angle $\alpha_0$;  regimes (i) and (ii) are separated at $ \alpha_{0, crit} \sim  \eta_0^{3/10}$. 
Resulting emission patterns show large variety: single or double peaked, and/or flat hat with sharp walls.  Multi-frequency  profiles - in optical and X-ray bands - can be used to get information about physical  (magnetic field strength, injection point) and geometrical  (dipolar angle and the line of sight) properties.}

\begin{document}

\maketitle  
\flushbottom


\section{Relativistic van Allen belts in astrophysical high energy sources}

A number of highly unusual compact binary systems show a presence of relativistic particles trapped in the magnetospheres   - astrophysical analogs of the terrestrial and Jovian  van Allen belts. These systems include the double pulsar companion PSR J0737-3039B,  low states of transitional millisecond pulsars like  {PSR J1023+0038}, as well as two exceptional magnetic  white dwarfs in  cataclysmic variables  AE Aqr and AR Sco. In all these systems, there is clear evidence of  magnetohydrodynamic interactions between stellar winds/accretion flow and the  \ms\ of the primary.  The  systems show non-thermal emission, extending  from radio to optical to X-rays,  and in some cases to   $\gamma$-rays. Emitting relativistic particles are likely accelerated during   magnetic reconnection events at the magnetospheric boundary. All these systems represent exotic paths/stages of binary evolution.

Van Allen radiation belts  in the  Earth \ms\  \citep[\eg][]{1961JGR....66.1321A} are  zones of energetic charged particles captured  from the solar wind (the Inner Van Allen radiation belt, consisting mainly of energetic protons, is the product of the decay of so-called ''albedo'' neutrons). They were one of the first  Solar wind  charged particles observed at the Earth; understanding them played a major role in the development of space science and plasma (astro)physics \citep{1950coel.book.....A}. Recent observations of  compact   astrophysical high energy sources   indicate a presence of similar population in  interacting   \NSs\ and {\WD}s binaries. These include a presence of dense plasma on the closed field lines of the double pulsar PSR J0737-3039B  \citep{2005ApJ...634.1223L}, particles producing high energy emission in highly asynchronous polars  AR~Sco and AE~Aqr \cite{2017NatAs...1E..29B,2017A&A...601L...7M,2020arXiv200411474L}, and low states  states of  transitional millisecond pulsars \citep{2014MNRAS.438.2105P,2015ApJ...807...33P}.

 \begin{figure}
 \includegraphics[width=.99\linewidth]{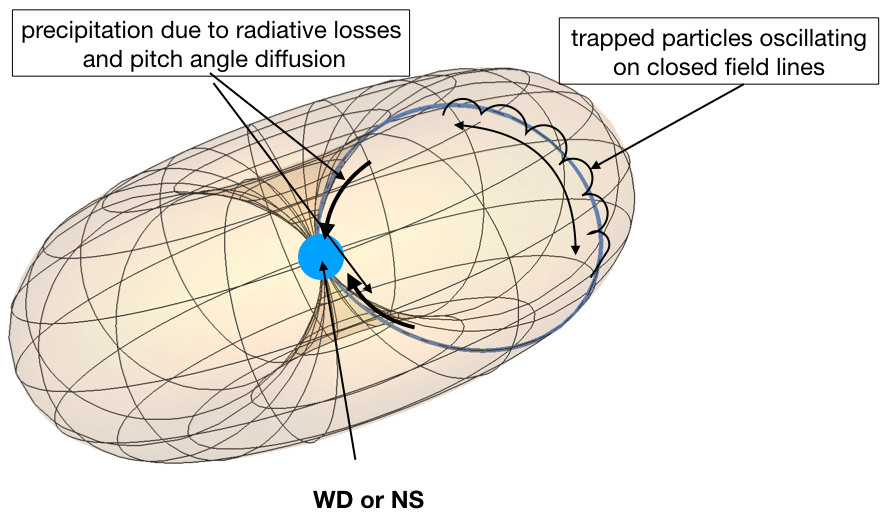}
\caption{\small Dynamic of relativistic particles trapped in the \mss\ of whites dwarfs/neutron  stars. Injected particles  experience adiabatic reflections between magnetic poles, lose energy to synchrotron radiation, and experience pitch angle scattering due to interaction with the turbulence.
  }
 \label{trapped-vanAllen}
 \end{figure} 
 
There are a number of principal differences  between terrestrial (and Jovian) van Allen belts and those expected in astrophysical compact objects  like {\NS}s and {\WD}s:   in astrophysical high energy sources radiative losses are/can be important (in the case of planetary \mss, the most important loss processes are collisions with neutrals).
In addition terrestrial plasma is mildly collisional: interaction of high energy particles with the neutral components is one of the dominant effects. In contrast, astrophysical plasma is highly collisionless.
The  plasma parameters in astrophysical sources are expected to be very different from the near space plasmas:  in {\NS}s plasma is expected to be magnetically dominated, and composed of electron-positron pairs. Both these effects change qualitatively the normal models of plasma and wave-particle interaction (\eg, there are no whistler modes in pair plasma).

  \section{Radiative and adiabatic forces for trapped particles}
  \label{radd}

  The particles comprising terrestrial van Allen belts are held trapped for a long time due to the conservation of first adiabatic invariant (magnetic bottling effect), Fig \ref{trapped-vanAllen}. The particles experience periodic motion between the magnetic poles, repulsed from the regions of strong \Bfs.
 The dominant non-adiabatic effects are pitch angle diffusion rates due to Coulomb collisions, non-elastic collisions with neutral  and resonant interactions with wave turbulence
\citep[\eg][]{1961JGR....66.1321A,1963P&SS...11..591D,1998JGR...103.2385A,1974ApJS...27..261C,2005Natur.437..227H}.

In relativistic \mss\ a new effect appears: radiative damping of cyclotron motion. Analysis of corresponding trajectories and emission patterns is the main goal of the present work.
Thus, the two main effects that control particle dynamics are conservation of the first adiabatic invariant and radiative losses.

Conservation of the first adiabatic invariant may in fact be expressed in terms of effective forces ${\bf F}_{ad}$, while radiative damping can be expressed in terms of radiative force ${\bf F}_{rad}$. We consider them next for  a relativistic  particle moving  an inhomogeneous magnetospheric  \Bf\  with \Lf\ $\gamma$ and pitch angle $\alpha$.

\subsection{Radiative damping}
\label{radiative}

Consider a particle moving in \Bf\ with pitch angle $\alpha$, so that velocities perpendicular and along the \Bf\ are
\ba &&
v_\perp = c \beta \sin \alpha
\nn &&
v_\parallel=  c \beta \cos \alpha
\ea
The radiation reaction force can be written as  \citep[][parag 76]{LLII}

\ba &&
F_{ \perp} = 
-\frac{2 \beta  B^2 e^4   \sin \alpha  \left(1+ \beta ^2 \gamma ^2 \sin ^2(\alpha
   )\right)}{3 c^4 m_e^2} = -{\beta   \sin \alpha   \left(1+ \beta ^2 \gamma ^2 \sin ^2\alpha \right)}  \frac{m_e c }{\tau_c} 
   \nn &&
   F_\parallel = -\frac{2 \beta ^3 B^2 \gamma ^2 e^4 \sin ^2\alpha  \cos (\alpha
   )}{3 c^4 m_e^2} = -{\beta ^3  \gamma ^2 \sin ^2\alpha  \cos \alpha } \frac{m_e c }{\tau_c} 
      \nn &&
     \tau_c = \frac{3 c^5 m_e^3}{2 B^2 e^4}
   \label{Frad}
   \ea

   Change of energy follows from
   \ba &&
   \partial_t (\gamma^2 - p^2) =0 \rightarrow  \partial _t  \gamma =  \v \cdot \partial _t {\bf p} =   \v \cdot {\bf F}
   \nn && 
     \frac{d\gamma }{dt}= \frac{{\bf F}_{rad}  \cdot {\bf v} }{m_e c^2}= - \frac{ \gamma ^2 \beta^2  \sin ^2\alpha }{\tau_c}
   \label{dgamma}
   \ea
   
    The momentum equation $\partial_t {\bf p} ={\bf F} _{rad}$ then gives
   \be
   \frac{d\alpha }{dt}=-\frac{ \sin (2 \alpha )}{2 \gamma  \tau_c}
      \label{dalpha}
   \ee
   Combing (\ref{dgamma}) and (\ref{dalpha}) we find
\be
   \frac{d\gamma }{d\alpha }=\gamma ^3 \beta^2  \tan \alpha 
\ee
with solution
\ba && 
\beta \mu = \beta_0 \mu_0
\nn &&
\mu = \cos \alpha
\label{betamu} 
   \ea
   where $\beta_0$ and $\alpha_0$ are initial velocity and pitch angle.
  Thus, parallel velocity $\beta \cos \alpha$ and parallel \Lf\ remains constant
  \be
   \gamma _{\parallel} = \frac{1}{\sqrt{1 -\beta^ 2 \mu^2}}={\rm const}
   \ee

Temporal evolution of \Lf\ is given by 
\be
\gamma  = \gamma_{\parallel} \coth \left( \frac{t}{\gamma_{\parallel}  \tau_c} +  {\rm arccoth} ( \gamma_0 /  \gamma_{\parallel} ) \right) 
\label{gammat}
\ee
This generalizes the known case with no parallel motion
\be
\gamma_\perp  = \coth\left( t/\tau_c + {\rm arccoth} \gamma_{\perp, 0} \right)
\label{gammat0}
\ee
Relation (\ref{gammat}) can also be derived from  (\ref{gammat0}) using Lorentz transformation for time and noting that $\gamma = \gamma_{\parallel}  \gamma_\perp$. 
Interestingly, though there is axial force, the axial velocity remains constant.

Evolution of pitch angle follows from (\ref{betamu}) and (\ref{gammat}):
\be
\mu = \frac{\beta _0 \mu _0}{\sqrt{1-\frac{\coth ^2\left(\frac{t}{\gamma _{\parallel,0} \tau
   _c}+\tanh ^{-1}\left(\frac{\gamma _0}{\gamma _{\parallel,0}}\right)\right)}{\gamma _{\parallel,0}^2}}}
   \ee
   As $t\to \infty$, $\mu \to 1$.

 After losing all the transverse momentum, $\alpha =0$, the particle is left with velocity 
   \be
   \beta_f= \beta_0 \cos \alpha _0
   \ee
   and  final energy 
\be
{ \gamma_f} = { \gamma _{\parallel} }
\ee

The additional radiation reaction force comes from curvature emission
\be
F_{\parallel, curv} = -\frac{2 \ \gamma ^4 e^2}{3 R_c^2}
\ee
where $R_c$ is the curvature radius.

For curvature emission to become important, the \Lf should be extremely high,
\be
\gamma \geq \sqrt{ c \om_B/R_c}
\ee
We assume that this regime is not achievable and neglect curvature emission.

\subsection{Adiabatic forces}

As a particle moves in inhomogeneous \Bf\ it conserves energy and the first adiabatic invariant. As a result, 
\be
\sin \alpha = \sqrt{\frac{B}{B_0} } \sin \alpha_0.
\ee
We can then formulate adiabatic force as
\ba &&
G_{\perp} = \pm \frac{\beta^2 \gamma m_e c^2}{4}   \frac{  ({\bf b} \nabla) B}{B} \sin 2 \alpha
\nn &&
G_\parallel = \mp \frac{\beta^2 \gamma m_e c^2}{2}   \frac{  ({\bf b} \nabla) B}{B} \sin ^2 \alpha
\ea
where ${\bf b} $ is unit vector along the \Bf and upper (lower) signs correspond to particle propagating towards (away from) the regions of increasing \Bf. 
Note that the expression $ \left( ({\bf b} \nabla) B \right) /B$ is {\it not } the curvature of field lines, which is  $  ({\bf b} \nabla){\bf b})$.

Since ${\bf G}\cdot {\bf v}=0$, $\gamma=const$ we can write the pitch angle evolution as 
\be
\partial_t \alpha = \pm \frac{  ({\bf b} \nabla) B}{2B} c \beta \sin \alpha. 
\ee

For particles propagating towards the star,  radiative  and adiabatic forces oppose each other in the evolution of the pitch angle. Which force ``wins'' depends in a subtle way on the parameters of the system and the injection location - this make this simple  system exceptional rich.

Qualitative, estimating  $ \left( ({\bf b} \nabla) B \right) /B \sim 1/L_B$, the ratio fo adiabatic and radiative forces
\ba &&
\frac{ G_{\perp}}{F_{ \perp}} \approx    \frac{\beta \gamma \cos \alpha}{ 1+ \beta^2 \gamma^2 \sin \alpha}   \times  \frac{ c \tau_c}{L_B}
\nn &&
\frac{ G_\parallel}{F_\parallel} \approx  \frac{1}{ \beta \gamma \cos \alpha}   \times  \frac{ c \tau_c}{L_B}
\ea

For not very small pitch angles, $\alpha \geq 1/\gamma$,
\be
\frac{ G}{F} \approx \frac{ c \tau_c}{\gamma L_B}
\ee

\subsection{Dipolar field}
Next we specify previous relations to   the case of dipolar fields.
A field line which  reaches $R_0$ at magnetic equator is given by
\ba &&
\frac{r}{R_0} = \sin^2\theta 
\nn &&
B = \sqrt{ \sin^2 \theta + 4 \cos^2 \theta} \left( \frac{R_0}{r}\right)^3 B_0=
\nn &&
={ \sqrt{4 -3 \frac{ r}{R_0}}}  \left( \frac{R_0}{r}\right)^3 B_0
\nn &&
ds = R_0 \sin\theta \sqrt{ \sin^2 \theta + 4 \cos \theta}  d \theta
\nn &&
\left( ({\bf b} \nabla) B \right) /B = -\frac{3 (27 \cos (\theta )+5 \cos (3 \theta ))}{\sqrt{2} r (3 \cos (2 \theta )+5)^{3/2}}
\label{eq:set1}
\ea
($ds $ is length along a given field line). We assumed that $R_0$ is the equatorial extension of  a field line, $\theta_0 = \pi/2$

In dipole field, a path along the field can be parametrized as
\be
ds = c \beta \cos \alpha \; dt
\label{eq:ds}
\ee
Using  Eqs.~(\ref{eq:set1}) and (\ref{eq:ds})  we can  find the evolution of the \Lf\ and the pitch angle
   \ba &&
   \frac{d\gamma }{dt}=-\frac{\beta ^2 \gamma ^2 \sin ^2\alpha  \csc ^{12}(\theta ) \left(\sin ^2(\theta )+4
   \cos ^2(\theta )\right)}{\tau _{c,0}}
   \nn && 
   \frac{d\alpha }{dt}= \pm \frac{3 \beta  c \sin \alpha  (27 \cos \theta +5 \cos (3
   \theta )) \csc ^2\theta }{2 \sqrt{2} R_0 (3 \cos (2 \theta )+5)^{3/2}}- \frac{\sin \alpha  \cos \alpha  \csc ^{12}(\theta ) \left(\sin ^2(\theta )+4 \cos ^2(\theta
   )\right)}{\gamma  \tau _{c,0}}   
   \nn &&
   \frac{d\theta }{dt}=\mp\frac{\beta  c \cos
   \alpha  \csc ^2\theta }{R_0 \sqrt{4 \cot ^2\theta +1}}
   \ea
where $\tau _{c,0}$ is defined in terms of the \Bf\ at $R_0$.

   Next we change to dimensionless quantities.
We normalize time to
   \be
   \tilde{t} = ( c t)/R_0
   \ee
   
   Another parameter is dimensionless cooling time at $R_0$, $ \tau _c$
   \be
\eta_c  =  \frac{R_0  }{\tau _{c,0} c } = \frac{2 B_0^2  e^4 R_0}{3 c^6 m_e^3}
\label{eq:etac}
   \ee
and change of parameters  $\tilde{r}=  r /R_0$.

We find
   \ba &&
   \frac{d\gamma}{d \tilde{t}} =-\eta_0  \gamma ^2  \beta^2 \sin ^2\alpha  (3 \cos (2 \theta )+5) \csc
   ^{12}(\theta )
   \nn && 
   \frac{d\alpha}{d \tilde{t}} = \pm \frac{3 \beta  \sin \alpha  (27 \cos (\theta )+5 \cos (3 \theta )) \csc ^2(\theta )}{2
   \sqrt{2} (3 \cos (2 \theta )+5)^{3/2}}
-\eta_0    \frac{2 \sin \alpha  \cos \alpha  \left(4 \cot ^2(\theta )+1\right) \csc ^{10}(\theta
   )}{\gamma } 
      \nn &&
   \frac{d\theta}{d \tilde{t}} =-\frac{ \beta  \cos (\alpha) }{ \sin^2 \theta \sqrt{5+3 \cos (2 \theta)}}
   \label{eq:dgdt}
   \ea
This is a set of equations for particle's velocity and \Lf, pitch anglee $\alpha$ and location in the dipole \mss\ $\theta$.

For $\eta_c  \geq 1$ (strongly cooling case)  a particle just loses most of the energy at the furthest point. The case of slow cooling,  $\eta_c  \leq 1$ is more interesting, we assume this regime below.


  Since we are interested in time-average state, the time differential can be replaced as   
\ba &&
dt= \mp \frac{\sec \alpha  \sqrt{4 R_0-3 r} }{2 \beta  c \sqrt{R_0-r}} dr
\label{eq:dtdr}
\ea

Substituting Eq.~(\ref{eq:dtdr}) to Eq.~(\ref{eq:dgdt}) we get 
   \ba&&
 \frac{d\gamma }{d \tilde{r}}= - \eta _c \frac{ \beta \gamma ^2  (4-3 \tilde{r})^{3/2} \sin \alpha  \tan
   \alpha }{  \sqrt{1-\tilde{r}} \tilde{r}^6}
   \nn &&
   \frac{d\alpha }{d\tilde{r}}= -  \eta _c \frac{ (4-3    \tilde{r})^{3/2} \sin \alpha }{\beta  \gamma  \sqrt{1-\tilde{r}} \tilde{r}^6}  \mp \frac{3 (8-5 \tilde{r})   \tan \alpha }{4 \tilde{r} (4-3 \tilde{r})}
   \label{eq:dgdr}
\ea
The system of equations  (\ref{eq:dgdr}) is our main  target: it describes the evolution of the energy and pitch angle for a particle experiencing synchrotron losses in dipolar \Bf. Initial conditions can be chosen as given \Lf\ $\gamma_0$ and pitch angle $\alpha_0$ at the magnetic equator $\tilde{r} =1$.

Since relations   (\ref{eq:dgdr})  are formally divergent near $\tilde{r} =1$, 
for numerical integration to start near $\tilde{r}=1$ one needs an analytical solution. At that point, there is no adiabatic term; we find
\ba &&
\gamma \approx  \gamma _0-{2 \gamma_0^2 \beta \eta_0  \sqrt{1-\tilde{r}_0} \sin \left(\alpha
   _0\right) \tan \left(\alpha _0\right)}
   \nn &&
   \alpha \approx \alpha _0-\frac{2 \eta _c
   \sqrt{1-\tilde{r}_0} \sin \left(\alpha _0\right)}{\beta  \gamma _0}
   \label{eq:gainit}
\ea

\section{Particle trajectories and synchrotron  emission patterns}

\subsection{Method}

The system   (\ref{eq:dgdr})   is non-Hamiltonian (non-energy conserving). This results in  a wide variety of behavior, typically  not scalable one from another.    As a result, to calculate the observed signal from trapped particles - the van Allen belts -  a fairly complicated procedure is required: (i) for each particle injected at a given point with a given pitch angle    equations of motion  should be integrated along dipolar \Bf\ lines; (ii) the emission pattern (Doppler-boosted  spectrum and direction in the observer frame) is then calculated; (iii) averaging over the initial distribution of pitch angles; (iv) averaging over the initial distribution of energies;  (v) for different relative orientations of a dipole with respect to the companion wind, the injection will occur at different points in the \ms.

Using standard Matlab procedures for stiff ordinary differential equations  \citep[ode15s from ][]{doi:10.1137/S1064827594276424}
with relative accuracy $10^{-7}$, we numerically solve the set of equations Eq.~(\ref{eq:dgdr}) with $\eta_0  = 1.25\times10^{-4}$
(this value corresponds to AR Sco WD in the model  of \cite{2020arXiv200411474L}) to calculate the evolution of Lorentz factor $\gamma_e$ and pitch  angle $\alpha$ along the particle trajectory.

We start particle motion from the magnetic dipole equatorial plane ($\tilde{r}=1$) along a given magnetic field line.  
If a particle reaches the mirror point ($\alpha=\pi/2$) or injection point ($\tilde{r}=1$), then we continue to calculate the trajectory until the particle loses 97\% of the initial energy (only 3\% of the initial energy left).

\subsection{Particle trajectories: bouncing, freezing and precipitation}
\begin{figure*}[]
\begin{center}
 \includegraphics[width=0.32\linewidth]{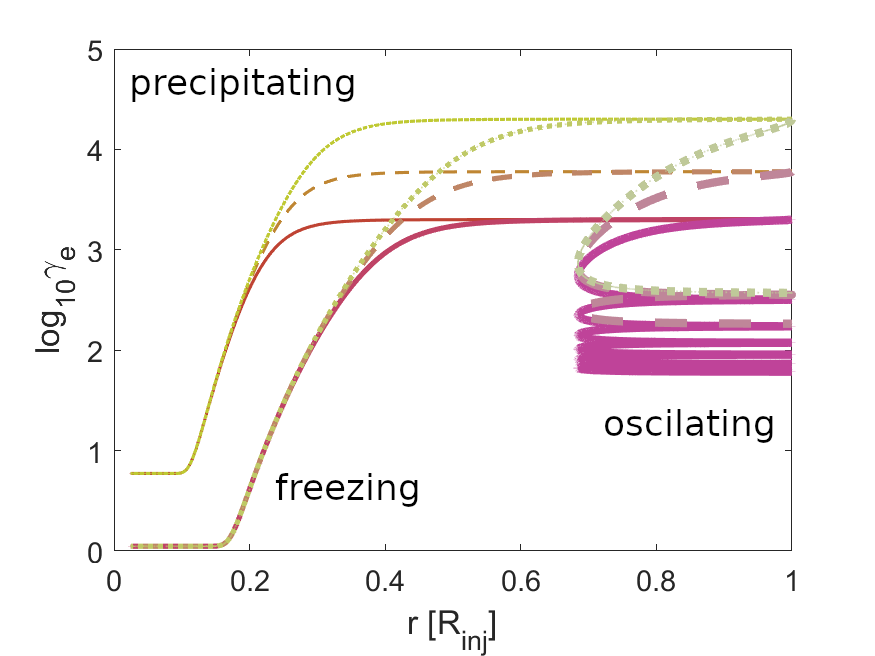}
\includegraphics[width=0.32\linewidth]{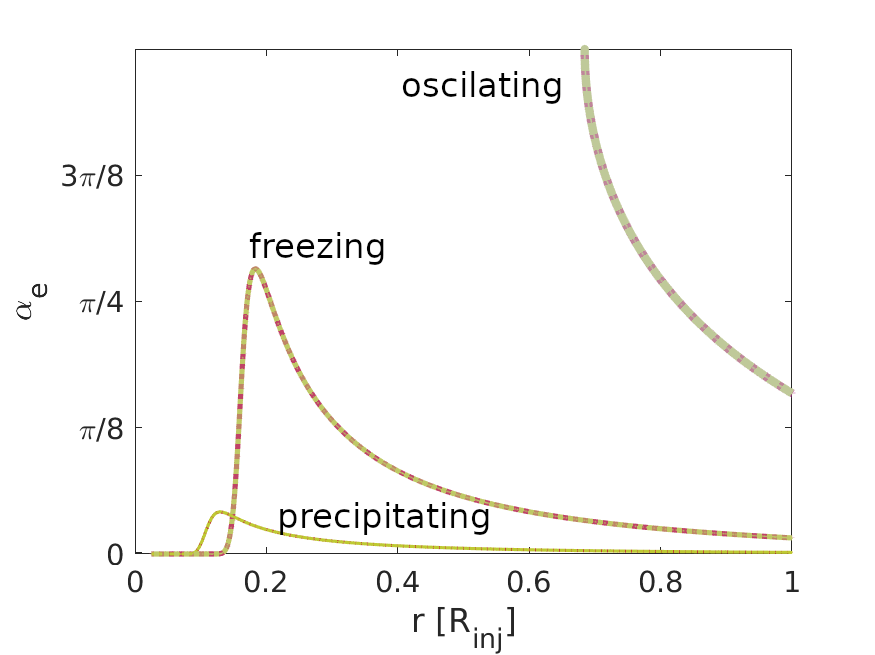}
 \includegraphics[width=0.32\linewidth]{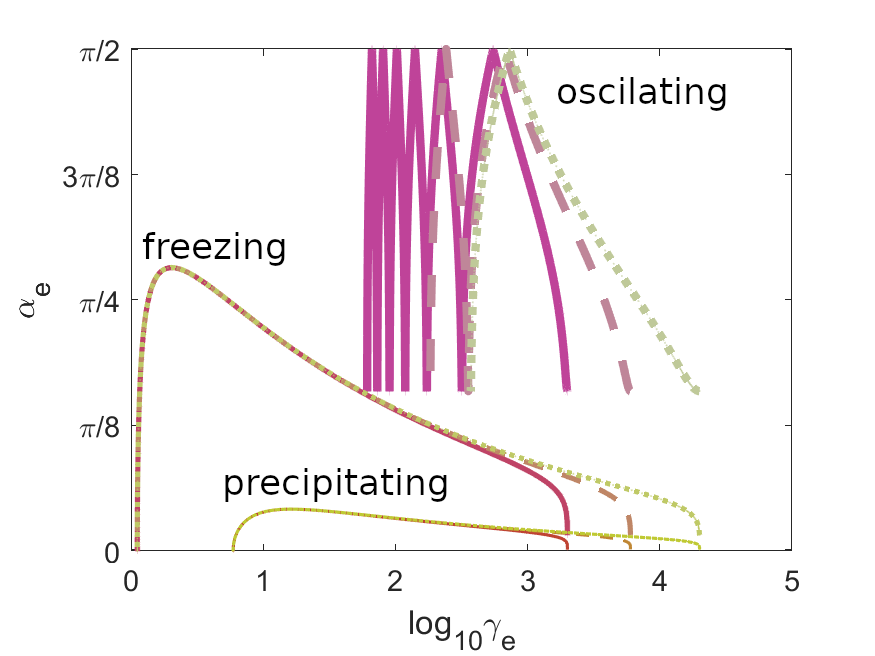}
   \end{center}
 \vspace{-12pt}
   \caption{ Left panel:  radial dependence of electron Lorentz factor $\gamma_e$, Center panel:  radial dependence of pitch angle $\alpha$,  Right Panel: 
trajectories of particles in the ``pitch angle  $\alpha$ -\Lf\  $\gamma_e$'' plane. There are three types of trajectories: ``oscillations'' , ``precipitation'', separated by  a ``freezing'' trajectory. Particles are injected at $\tilde{r}=1$ with different {\Lf}s and different pitch angles. 
The thin lines corresponds to $\alpha_{e,0} = 0.005$, the medium $\alpha_{e,0} = 0.05$ and thick ones $\alpha_{e,0} = 0.5$. The 
solid lines corresponds to $\gamma_{e,0} = 2\times10^3$, the dashed $\gamma_{e,0} =  6\times10^3 $ and dotted  $\gamma_{e,0} =  2\times10^4$. 
Particles with high initial pitch angle ($\gtrsim 0.1$) oscillate between reflection points with slowly decreasing energy. For sufficiently small pitch angles ($\sim 0.05$) due to high radiative losses  particle propagate deep into the \ms, and quickly lose both parallel and perpendicular momentum. For even lower pitch angles ($\lesssim 0.01$) particle lose all the transverse energy and fall onto the star. 
 }
 \label{fig:gar}
\end{figure*}

 There are  three  types of particle trajectories: (i) bouncing: particle can experience multiple  bounces between magnetic bottles near  the pole (this is close to the conventional regime when the radaitive losses are weak; (ii) precipitation:  particles lose their transverse motion and fall onto the star; (iii) freezing  particles lose most of their transverse  and  nearly all the parallel momentum,  Fig. \ref{fig:gar}.

In case when both radiative damping and adiabatic forces are present, there is an interesting case which we call the  ``freezing'' trajectory. Qualitatively, as a particle propagates towards regions of higher \Bf, the adiabatic force acts to slow down parallel motion, and increase the transverse velocity. As  result, a  particle spends a lot of time near the reflection points, with large transverse momenta.  Particles then  experience large radiative losses; this is the  ``freezing'' trajectory. 

Let us approximate $ ({\bf b} \nabla) B/{B} \sim 1/L_B$.
A particle spends 
\be
   \tau _B \sim L_B / c
   \ee
   near the reflection point.  This is a region with largest \Bf\ allowed by the first adiabatic invariant: a given particle  spends a lot of time in the region with highest \Bf. 
   If radiative decay time at the reflection point is shorter that $ \tau _B $ then a particle  exponentially lose its energy and will not bounce back.

   Estimating the reflection point as $ \tilde{r} \sim \sin ^{2/3} \alpha_0$, the freezing case corresponds to balancing two terms in the expression for $\alpha$ in (\ref{eq:dgdr}):
   \be
 \alpha_0 \sim  \eta_c ^{3/10}
 \label{alpha0}
   \ee
   For larger initial pitch angles particles experience bouncing trajectories, for smaller ones they precipitate.

\begin{figure*}[]
\includegraphics[width=0.49\linewidth]{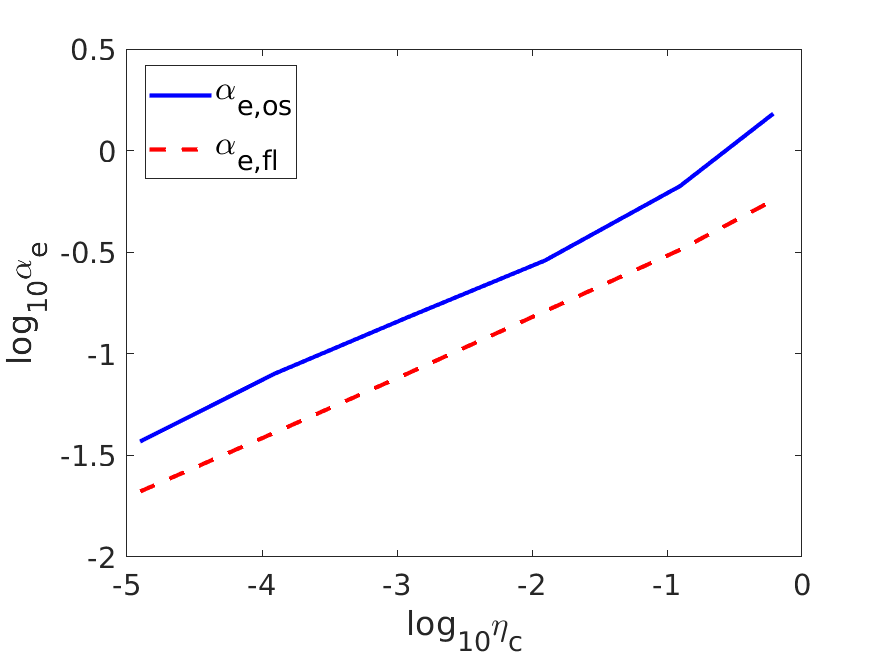}
\includegraphics[width=0.49\linewidth]{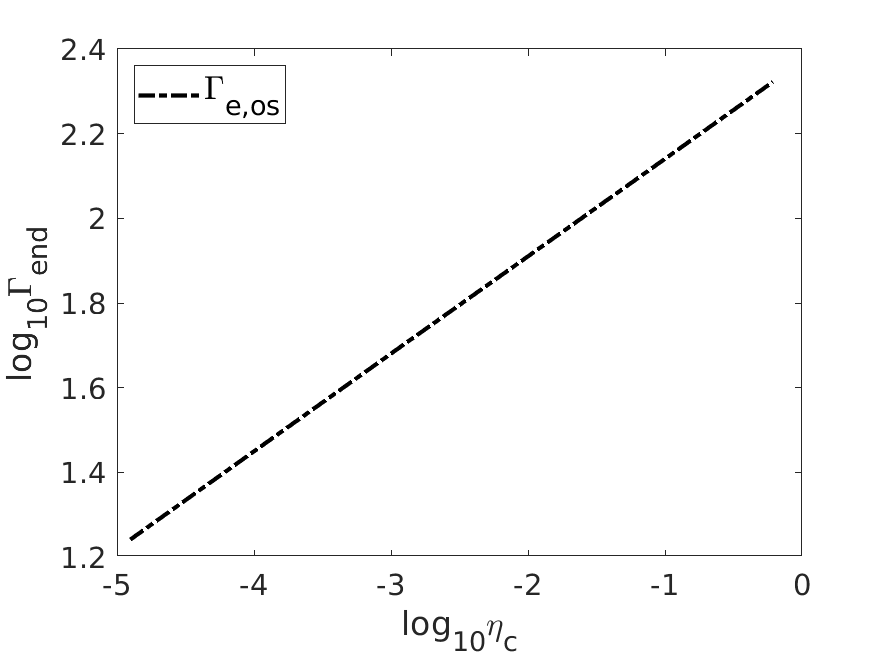}
\caption{We plot data based on the Table~\ref{tab:models}. On the left panel, the maximal pitch angle of falling/``precipitation'' trajectories depends on the $\eta_0 $ parameter are presented by dashed line, the maximal pitch angle of the ``frozen'' trajectory  presented by solid line. On the right panel is shown the logarithm of the maximal Lorentz factor of the falling particles by dot-dashed line.  }
 \label{fig:etaaG}
\end{figure*}
 
To  illustrate  particle's trajectories,  we present nine solutions corresponding  to  a combination of initial pitch angle $\alpha_{e,0} = 0.005,\; 0.05,\; 0.5$  and $\gamma_{e,0} = 2\times10^3,\;   6\times10^3,\; 2\times10^4$.  On the Fig.~\ref{fig:gar} we present radial dependence of the  pitch angle (right)  and Lorentz factor (center panel). To make trajectories more clear, we plot dependence of pitch angle from Lorentz factor. We can see three definite classes of trajectories: 
1) large initial pitch angle leads to formation of ``oscillation'' trajectory;  2) the intermediate pitch angle leads to initial grow of pitch angle (up to value order of $\alpha_{e}\sim1$) and cooling of Lorentz factor to 1 with collapse of pitch angle to 0, so called ``frozen'' trajectories; 3) the small pitch angles after moderate grow leads to collapse of pitch angle to 0 with $\gamma_e>1$, so called ``precipitation'' or falling trajectories.

The analysis of the trajectories shows that  the characteristic trajectory is determined by initial pitch angle. The particles with $\gamma_e > 10^2$ quickly lose energy and the latter on trajectories looks similar.

We investigated the particle trajectories for a wide variety of the $\eta_0 $ parameters. Three types of trajectories can be seen for all investigated cases $\eta_0 <1$. Oscillating  trajectories are more probable for  $\eta_0 \ll1$, on the other hand, frozen trajectories and  precipitation trajectories are to keep its relative width independently of parameter $\eta_0 $. Summary of these results are  presented in  Table~\ref{tab:models} and Fig.~\ref{fig:etaaG}. The critical pitch angle, which separates falling/``precipitation'' trajectory  can be estimated with accuracy better 1\% as 
\be
\alpha_{e,fl} = 0.589\eta_{c}^{0.30},
\ee
consistent with (\ref{alpha0}). 
 For the ``frozen'' trajectory, the pitch angle should be less when  
 \be 
 \alpha_{e,os} = 1.079\eta_{c}^{0.30}.
 \ee
  This expression works with accuracy better 1\% for $\eta<0.1$.
The maximal Lorentz factor of the falling particles can be estimated as $$\Gamma_{end}=234\eta_{c}^{0.23}.$$

\begin{table}[]
\centering
\caption{The critical parameters of the trajectories. }
\label{tab:models}
\begin{tabular}{cccc}
\hline
 $\eta_{ c}$      &   $\alpha_{e,os}$  &   $\alpha_{e,fl}$ &  $\gamma_{e,end,max}$ \\
\hline
&&&\\[-5pt]
 \quad $1.25\times10^{-5}$ &  \quad 0.037 &\quad 0.021 & \qquad 17.42 \\
 \quad $1.25\times10^{-4}$ &  \quad 0.080 &\quad 0.041 & \qquad 29.65 \\
 \quad $1.25\times10^{-3}$ &  \quad 0.153 &\quad 0.081 & \qquad 50.44 \\
 \quad $1.25\times10^{-2}$ &  \quad 0.288 &\quad 0.162 & \qquad 85.60 \\
 \quad $1.25\times10^{-1}$ &  \quad 0.669 &\quad 0.325 & \qquad 145.5 \\
 \quad $ 0.613$            &  \quad 1.52  &\quad 0.57  & \qquad 210 \\
\hline
\end{tabular}
\caption{
For dimensionless cooling time parameter $\eta_0 $ (see Eq.~\ref{eq:etac}) and $\gamma_e \gg 10^2$ we presented critical initial pitch angles, which separated different kinds of trajectories: 1) oscillation -- $\alpha_{e,0}>\alpha_{e,os}$; 2) freezing -- $\alpha_{e,os}\ge\alpha_{e,0}>\alpha_{e,fl}$; 3) precipitation -- $\alpha_{e,fl}\ge\alpha_{e,0}$. The maximum Lorentz factor of falling electrons is presented as $\gamma_{e,end,max}$ ($\gamma_{e,fl,max}$ measured at $\alpha_{e.0} = 0.00061$). 
}
\end{table}

\subsection{Injection procedure and emissivity}

For the location of injected particles, we chose a particular radius and inject at various polar angles. Injection at various polar angles allows us to mimic various accretion geometries and orientations of the magnetic axis of the accelerator. For the fully  aligned case, the injection is expected at magnetic equator, $\theta= \pi/2$, Figs. \ref{fig:galum}- \ref{fig:lummaps}. In addition, we investigated the case of $\theta =\pi/8, \, \pi/6,\, \pi/4,\, \pi/3$ (Fig. \ref{fig:lummapsbig}). 

\begin{figure}[]
\includegraphics[width=0.95\linewidth]{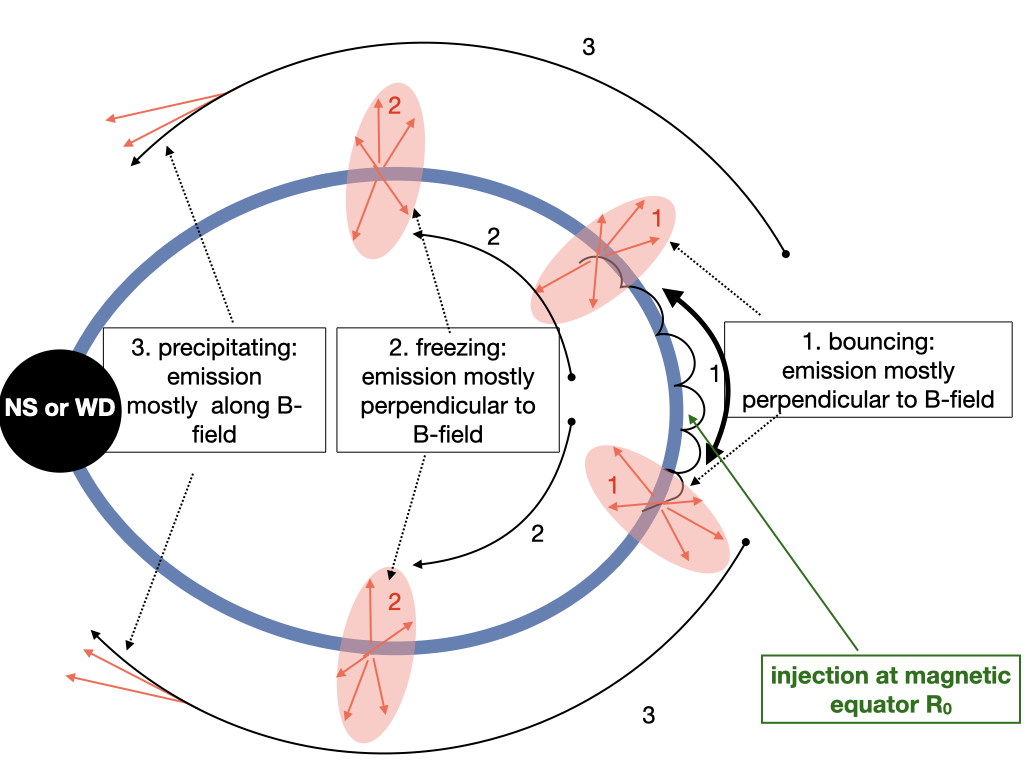}
\caption{Cartoon of different particles' trajectories and emission patterns. In the slow cooling case (1) particles mostly bounce between reflection points producing emission mostly near the reflection point orthogonally to the \Bf. 
For the  freezing case (2)  particles lose nearly all the energy in one go and fall to the star with small energies;  emission is produced also mostly orthogonally to the \Bf. 
 In the case of precipitation (3) particle keep their relativistic parallel momenta, so that emission is beamed along \Bf. Qualitatively, for our parameters,  it is trajectory 1 that dominate  overall  emission (more particles are in this regime).}
 \label{ftrapped-vanAllen1-2}
\end{figure}

To construct emission pasterns, we inject an ensemble of particles with uniform distribution in initial velocity direction and spectrum  $dN/d\gamma_e\propto \gamma_e^{-2}$. The emission patterns are the intensity of the photon emission depends on two viewing angles relative to the WD center and magnetic axis $\theta$ and $\phi$. The $\theta$ and $\phi$ are altitude and azimuthal angles, respectively. For the particle emission point $(\phi,\theta) = (0,\pi/2)$ emission patterns in optics and X-ray are presented on the Fig.~\ref{fig:lummaps}, here we change $\eta_0 $ parameter from $1.25\times10^{-3}$ (top raw) till $1.25\times10^{-5}$ (bottom raw). As we can see for $\eta_0 \ge 10^{-4}$ the emission pattern is simple, equatorial zone dominates in the emission and the light curve will be nearly constant if rotation axis is parallel to the magnetic one. For smaller values of the $\eta_0 $ parameter we can see two distinct peaks which will form two maximum on the light curves as in optics as in X-ray. In general, light curves as in optics as in X-ray look very similar.

The synchrotron power  of each particle can be  calculated as $L_{syn} = d\gamma_e/dt$. 
To calculate  the synchrotron,  for a given \Lf\ and pitch angle, we use the most probable frequency  of 
\be
\nu = 0.29\frac{3   B e}{4\pi m_e c}\gamma_e^2 \sin^2\alpha.
\label{eq:numax}
\ee

\begin{figure}[]
  \includegraphics[width=0.99\linewidth]{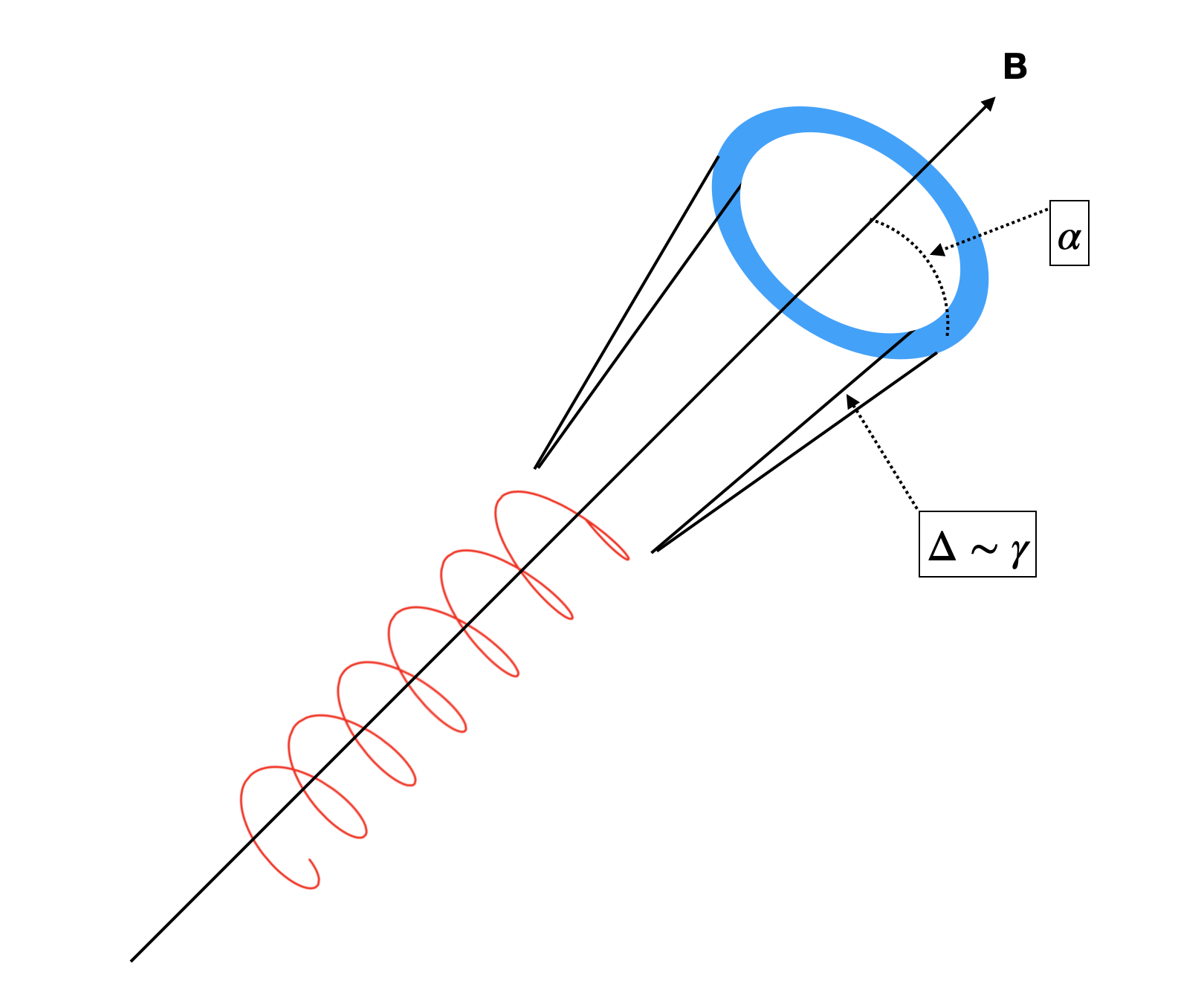}
   \caption{Emission pattern
 }
 \label{emission-pattern}
\end {figure}

At each point, emission is beamed into a  conical diagram with opening angle $ \Delta \sim 1/\gamma$ around the pitch angle $\alpha$, Fig. \ref{emission-pattern}. When motion becomes mildly relativistic, $\gamma_e \sim $ few and the emission diagram becomes wide, the synchrotron emissivity is small.

Thus, for each particle injected with a  given pitch angle $\alpha_0$ and given \Lf\ $\gamma_e$,  we follow its evolution by solving system (\ref{eq:dgdr}). Then at each point we calculated the intensity and direction of synchrotron emission,  and project it onto the sphere of the sky.

\subsection{Emission patterns and spectra}
 
\begin{figure}[]
  \includegraphics[width=0.99\linewidth]{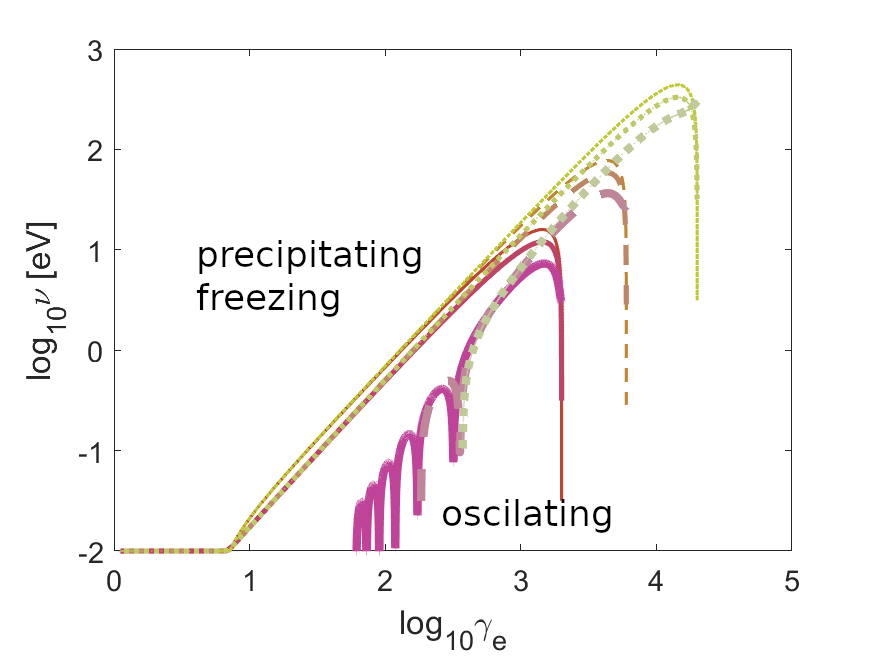}
   \caption{ Trajectories of particles in the \ms\ plotted in the   ``frequency -- Lorentz factor" plane, equatorial injection.  {\Lf}s (increasing to the right) and different pitch angles  (increasing from top to bottom in each set).
    The thin lines corresponds to $\alpha_{e,0} = 0.005$, the medium $\alpha_{e,0} = 0.05$ and thick ones $\alpha_{e,0} = 0.5$. The 
    solid lines corresponds to $\gamma_{e,0} = 2\times10^3$, the dashed $\gamma_{e,0} =  6\times10^3 $ and dotted  $\gamma_{e,0} =  2\times10^4$.    
 }
 \label{fig:trajectories1}
\end {figure}

Next, we calculate the light curves and spectra of synchrotron emission. First, in Fig.~\ref{fig:trajectories1} we show particles' trajectories in the emitted frequency - \Lf\ plane for equatorial injection.
Particles first  start at the highest {\Lf}s, but in a low \Bf, so that emitted frequency is relatively low. As they propagate into stronger \Bf,  and increasing their pitch angle, the emitted frequency  increases.

Spectral properties of the resulting emission 
are  presented in   Fig.~\ref{fig:trajectories1}. The Lorentz factor is the main factor, but in general the dependence is non-trivial, as less energetic particles propagating deep into the \ms produce higher energy photons, while higher energy particles gyrating at a large distance produce lower frequencies.  For example,  for relatively large pitch angles and large Lorentz factors the emission frequency drops down due to particle oscillations at relatively large distances (see Fig.~\ref{fig:gar}) and smaller magnetic field strength.
For the parameters characteristic to  system AR~Sco at higher energies, in optics  emission is produced by electrons with energy about 1~GeV and X-rays are formed by electrons with energy larger 20~GeV. Interesting that the largest part of emission is produced with high pitch angles $\alpha_e>\pi/4$, but maximum photon energy vary on factor $\sim1.5$ for difference in initial pitch angles from 0.005 till 1.

\begin{figure*}[]
\includegraphics[width=0.49\linewidth]{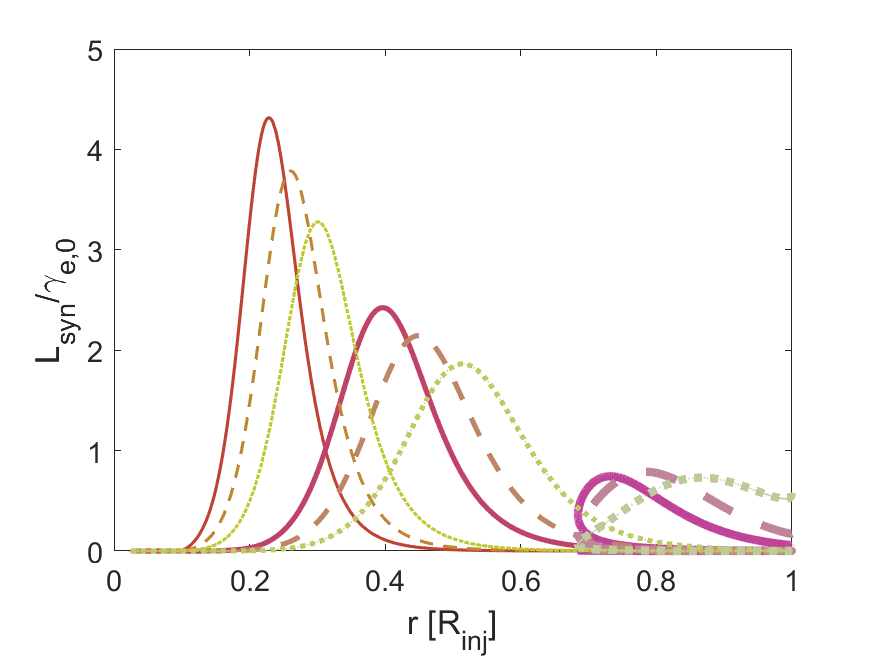}
\includegraphics[width=0.49\linewidth]{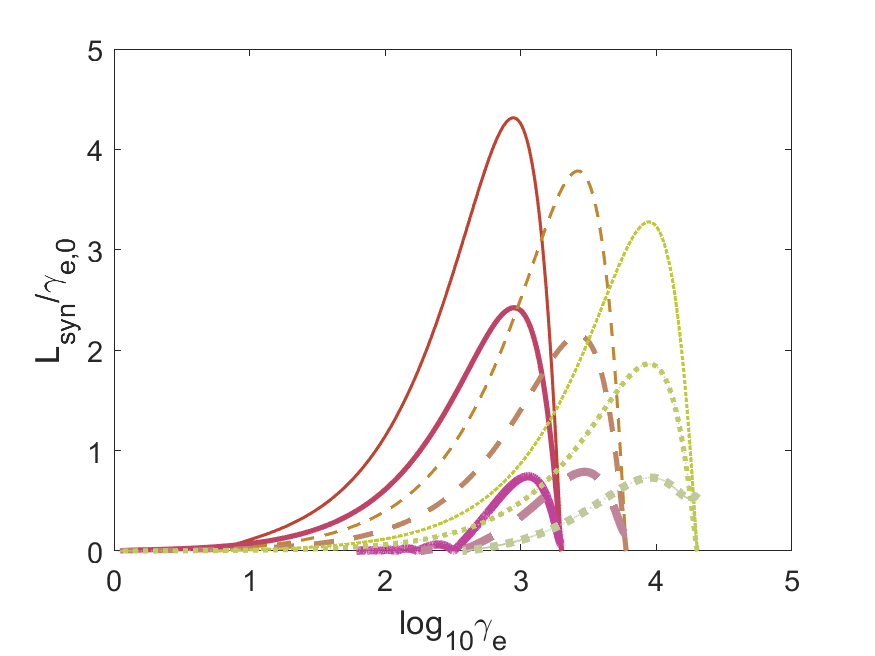}
\includegraphics[width=0.49\linewidth]{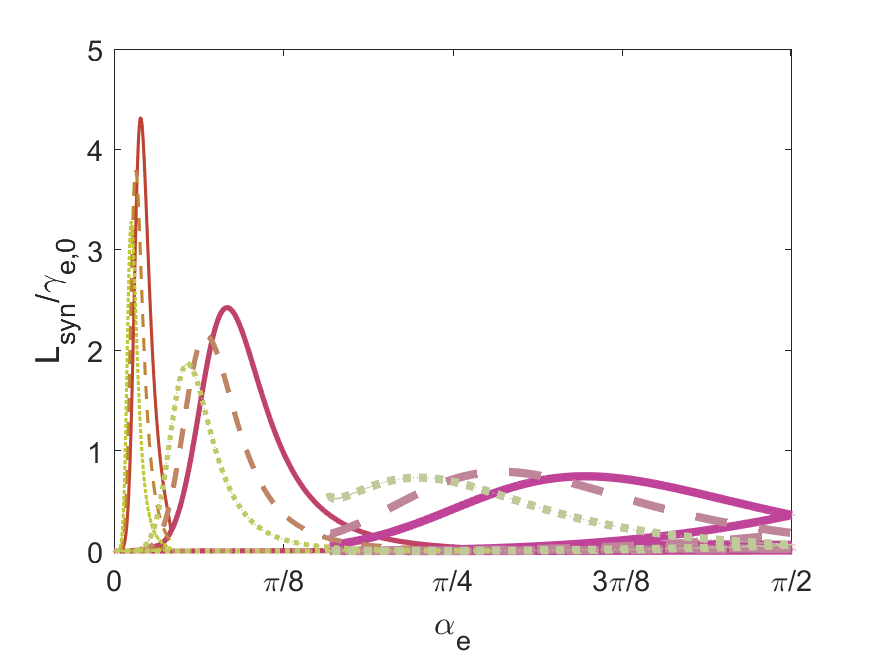}
\includegraphics[width=0.49\linewidth]{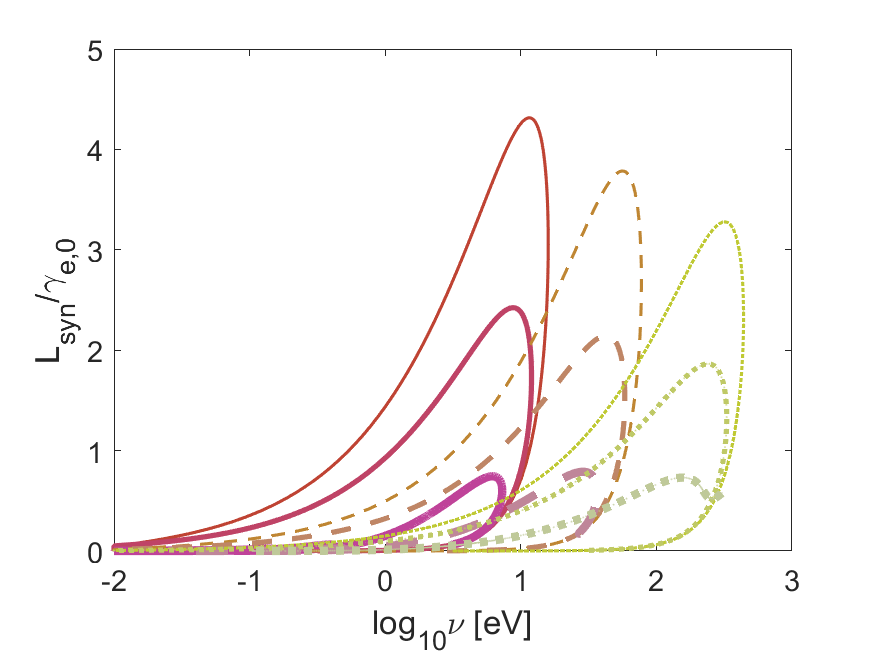}
\caption{Equatorial injection. The synchrotron luminosity normalized on initial Lorentz factor depends on radius (top left),  electron Lorentz factor $\gamma_e$ (top right),  pitch angle $\alpha$ (bottom left) and 
peak flux frequency of synchrotron photons $\nu$ (bottom right) are  presented. The thin lines corresponds to $\alpha_{e,0} = 0.005$, the medium $\alpha_{e,0} = 0.05$ and thick ones $\alpha_{e,0} = 0.5$. The 
solid lines corresponds to $\gamma_{e,0} = 2\times10^3$, the dashed $\gamma_{e,0} =  6\times10^3 $ and dotted  $\gamma_{e,0} =  2\times10^4$.  }
 \label{fig:galum}
\end{figure*}

On the Fig.~\ref{fig:galum} we show  dependence of normalized synchrotron  luminosity depends on radius (left top), Lorentz factor (right top), pitch angle(left bottom) and emission frequency (right bottom) for $\eta_0 =1.25\times10^{-5}$. The luminosity is a strong function as a Lorentz factor  as a pitch angle. The higher Lorentz factor leads to grow of luminosity. The small pitch angle allows particle travel to high magnetic field region before significant loses of the initial Lorentz factor and the growth of the pitch angle in the region with high magnetic field strength leads to more bright and hard emission. Even for small value of the $\eta_0 $ parameter, the main energy loss process for particle's majority take place at relatively large heights $0.5<r<0.9$.

\begin{figure*}[]
\includegraphics[width=0.49\linewidth]{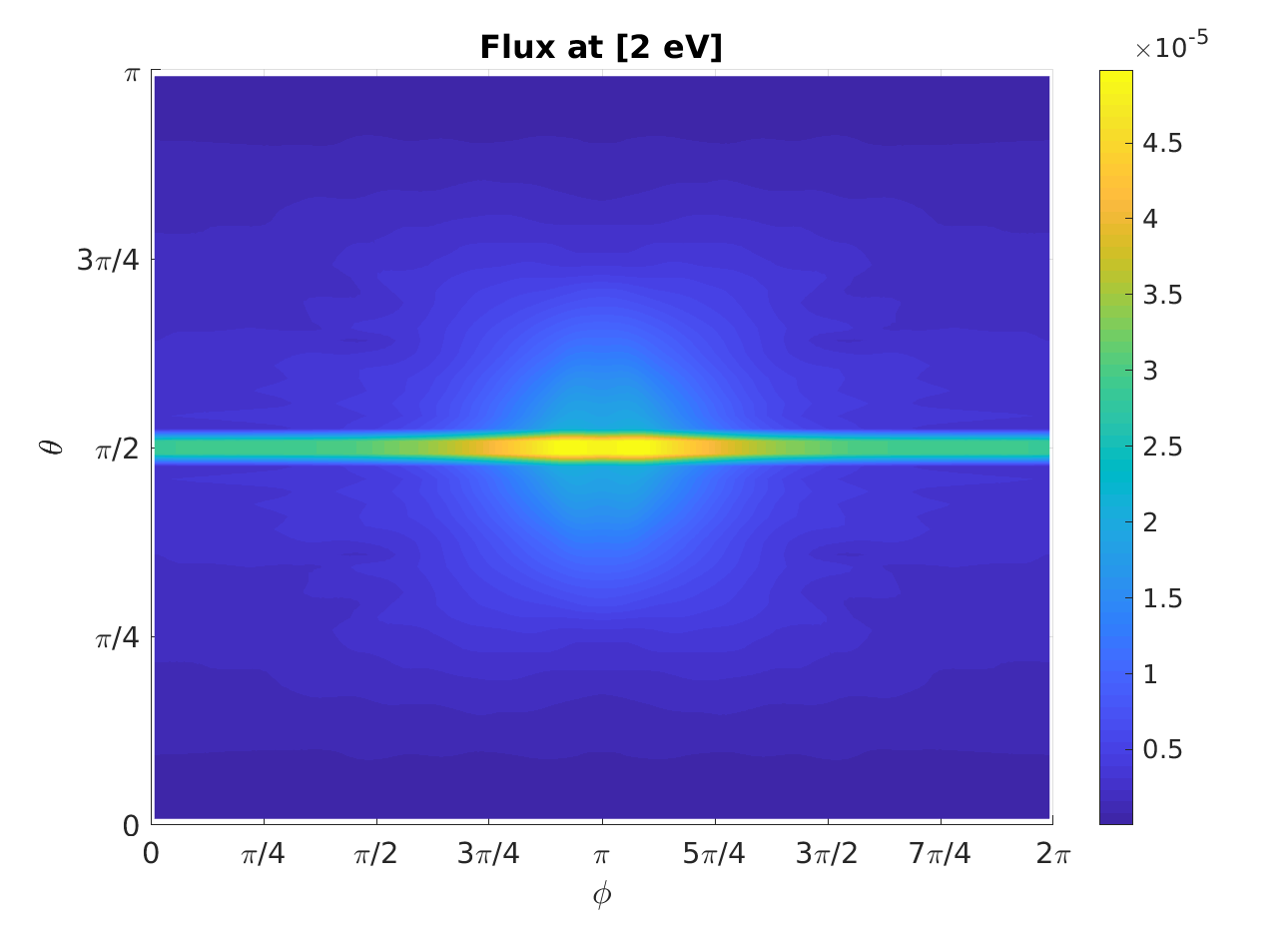}
\includegraphics[width=0.49\linewidth]{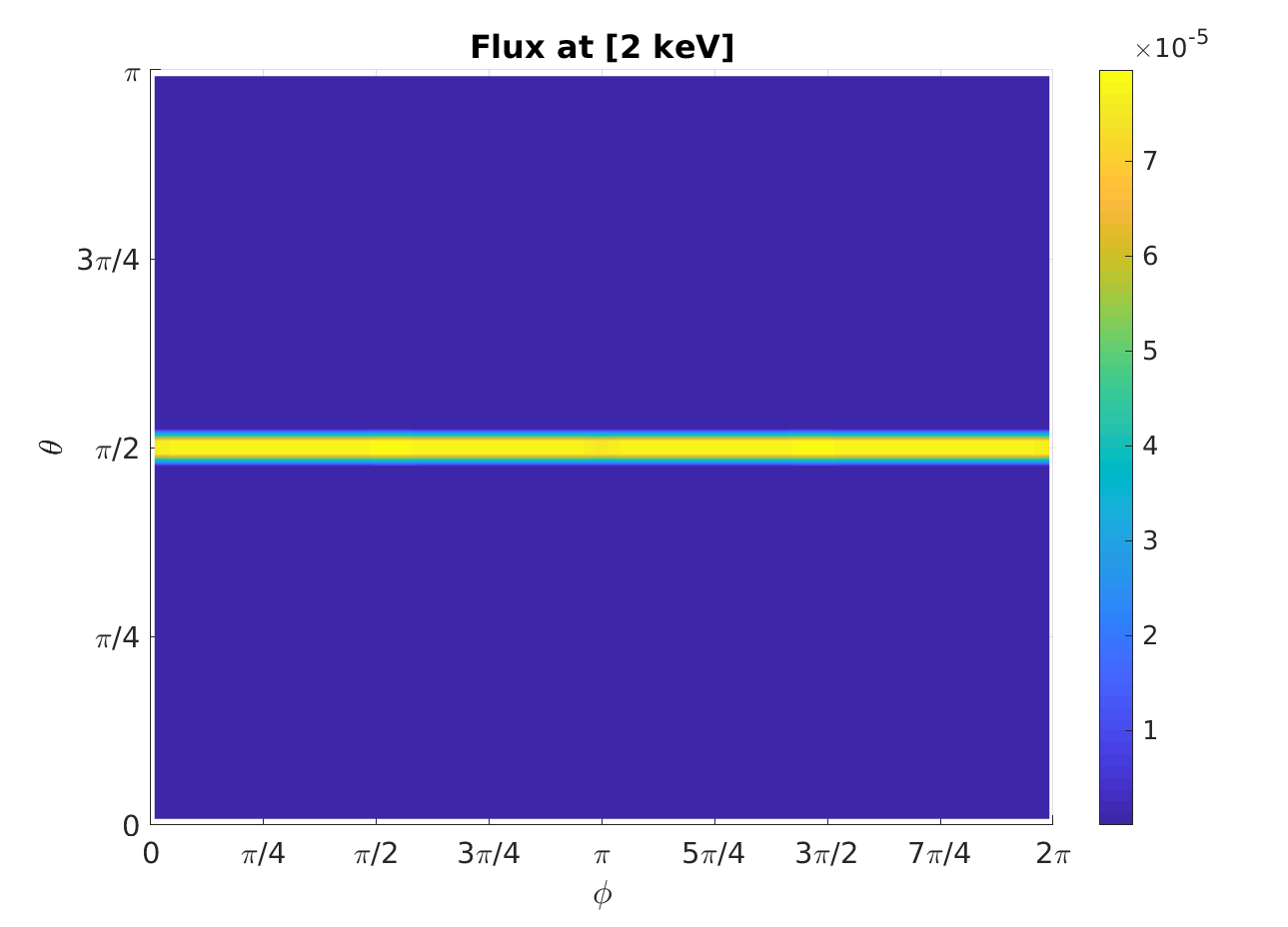}
\includegraphics[width=0.49\linewidth]{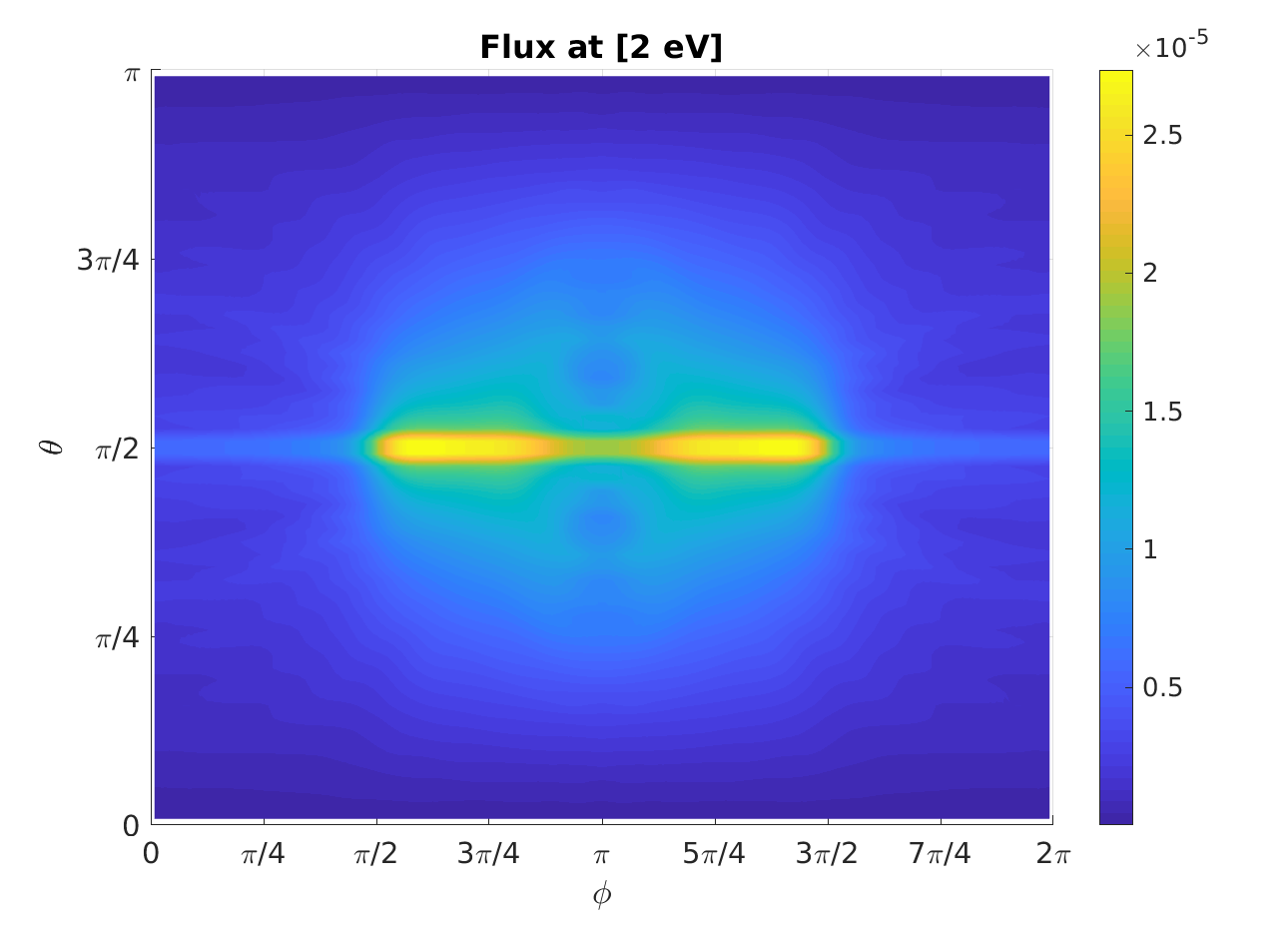}
\includegraphics[width=0.49\linewidth]{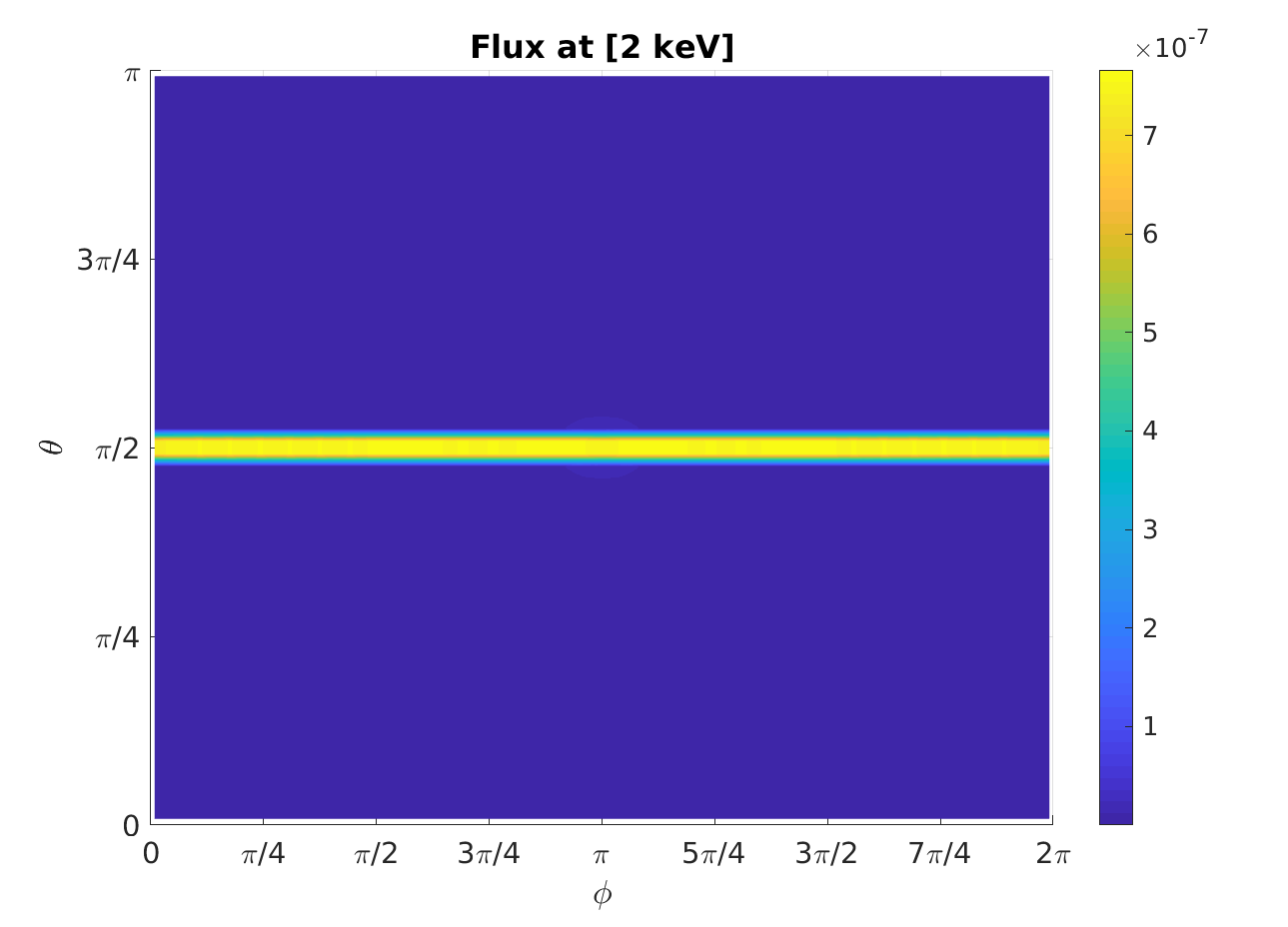}
\includegraphics[width=0.49\linewidth]{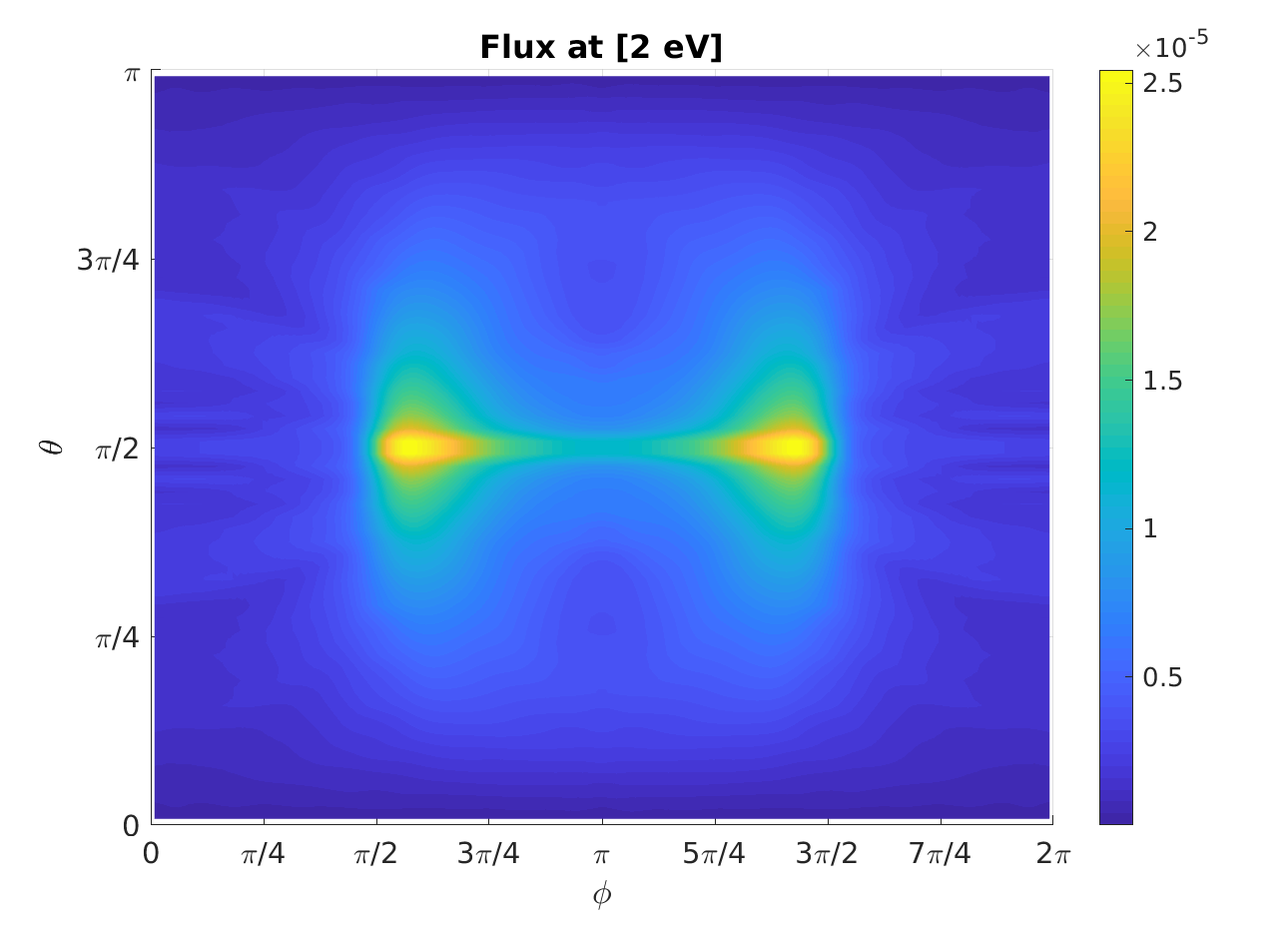}
\includegraphics[width=0.49\linewidth]{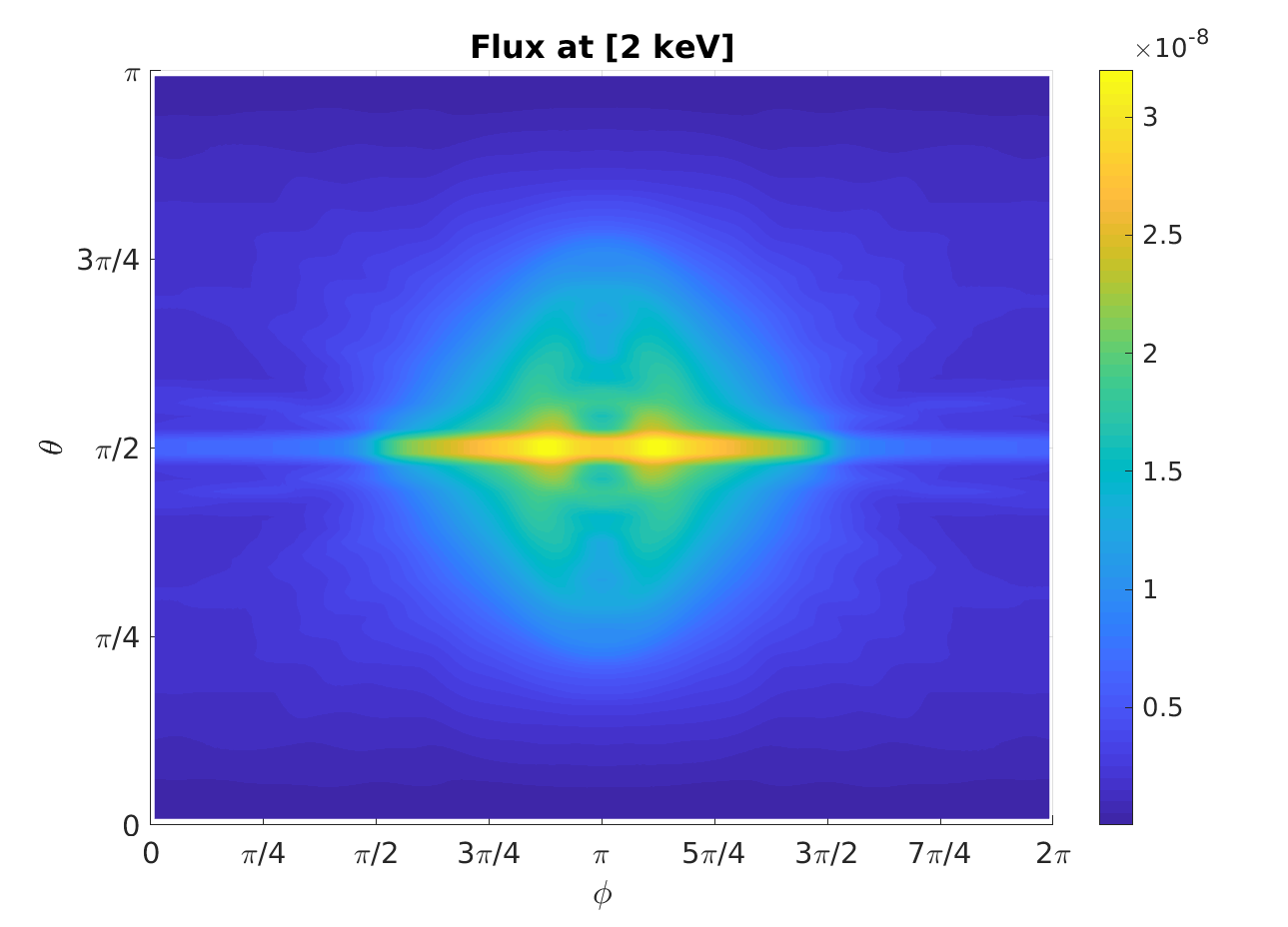}
\caption{
The synchrotron luminosity maps in optical (left) and in X-ray (right) energy band depends on viewing angle ($\theta,\phi$). Here $\theta = 0$ in direction of magnetic axis, $\phi = 0$ in direction of magnetic field line plane, and the injection point $(\theta,\phi)=(\pi/2,0)$.
The $\eta_0 $ was varied in the range $1.25\times10^{-3}$ top row, $1.25\times10^{-4}$ middle row and $1.25\times10^{-5}$ bottom row. }
 \label{fig:lummaps}
\end{figure*}

In Fig. \ref{fig:lummaps} we plot sky  emission maps at different photon energies and different radiation (cooling)  parameter  $\eta_0$. Particles are injected only at fixed azimuthal angle $\phi=0$. 
For strong cooling, $\eta \geq 10^{-4}$, particles lose most of the energy near the initial equatorial plane, while producing emission mostly in the equatorial plane. Qualitatively, top and middle rows correspond to trajectory 1 in Fig. \ref{ftrapped-vanAllen1-2}.

For weaker cooling, smaller parameters $\eta$ (bottom row in Fig. \ref{fig:lummaps}, the picture change: a particle has time to move from the equatorial plane towards the magnetic axis. As a result, emission is produced into a wider area of the sky. Yet the points $\phi = \pi/2 $ $3\pi/2$ are special: a particle produces emission in those directions along the whole trajectory, so that the emission accumulates towards this direction. 

\begin{figure*}[]
\includegraphics[width=0.45\linewidth]{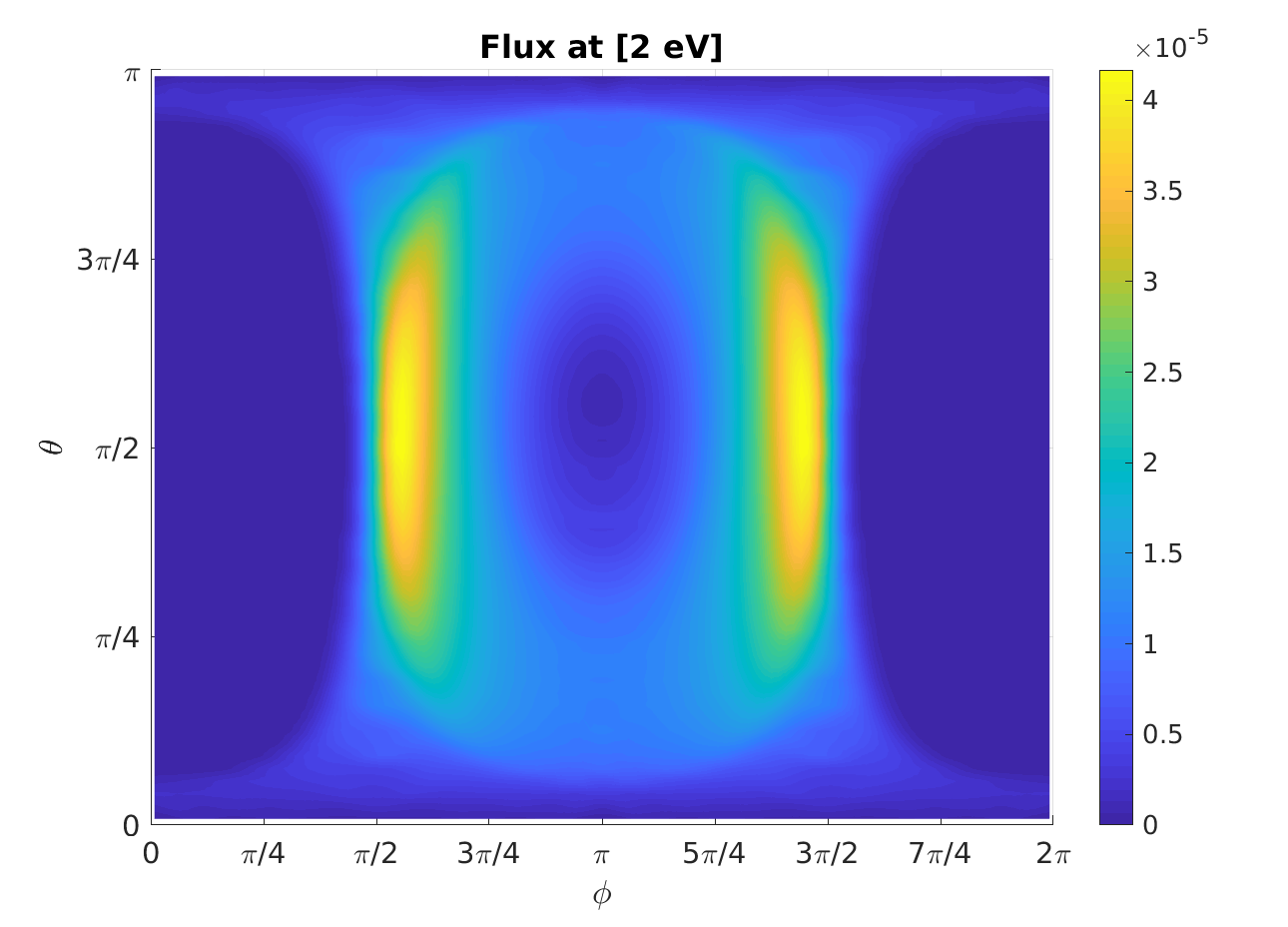}
\includegraphics[width=0.45\linewidth]{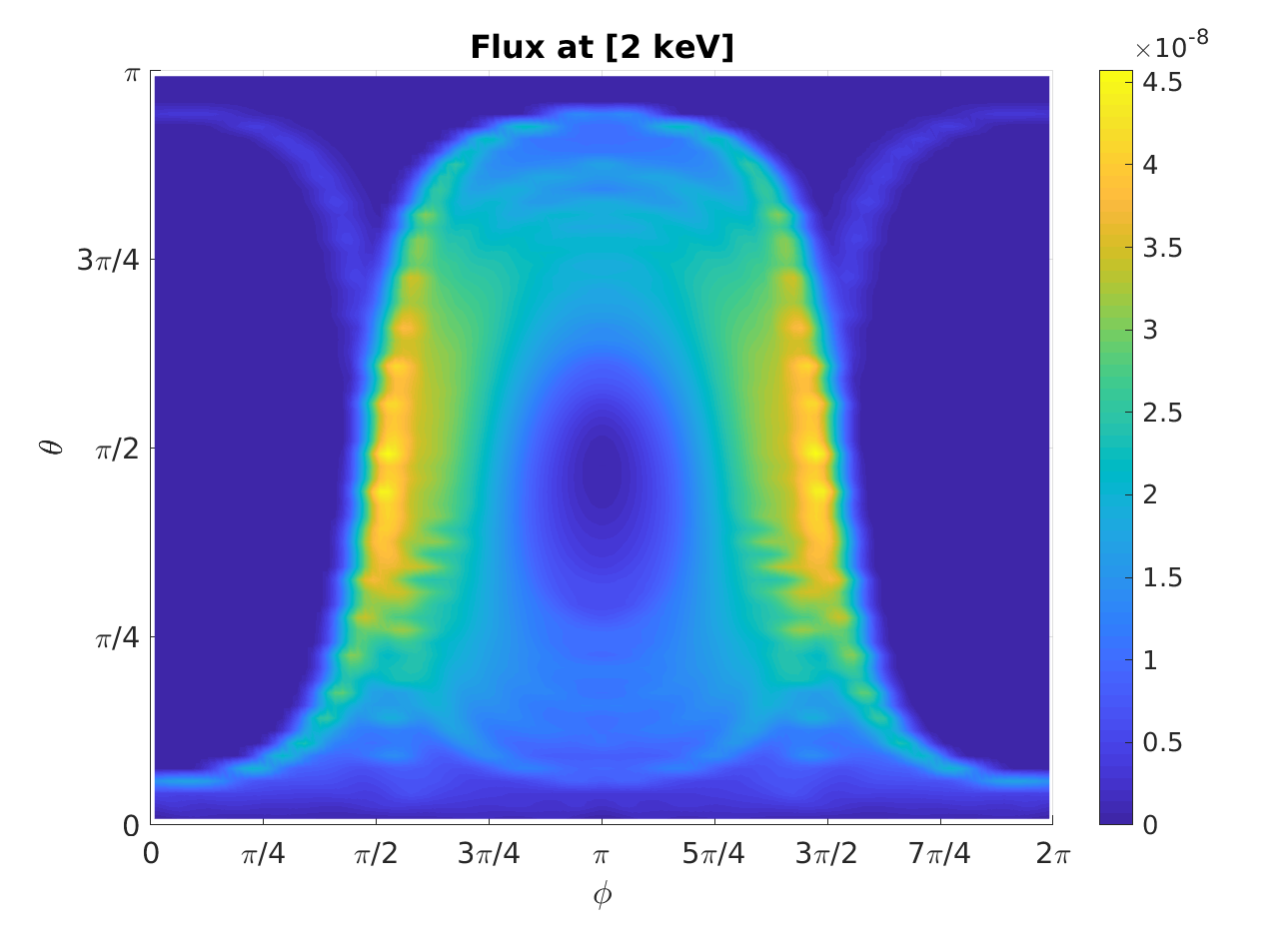}
\includegraphics[width=0.45\linewidth]{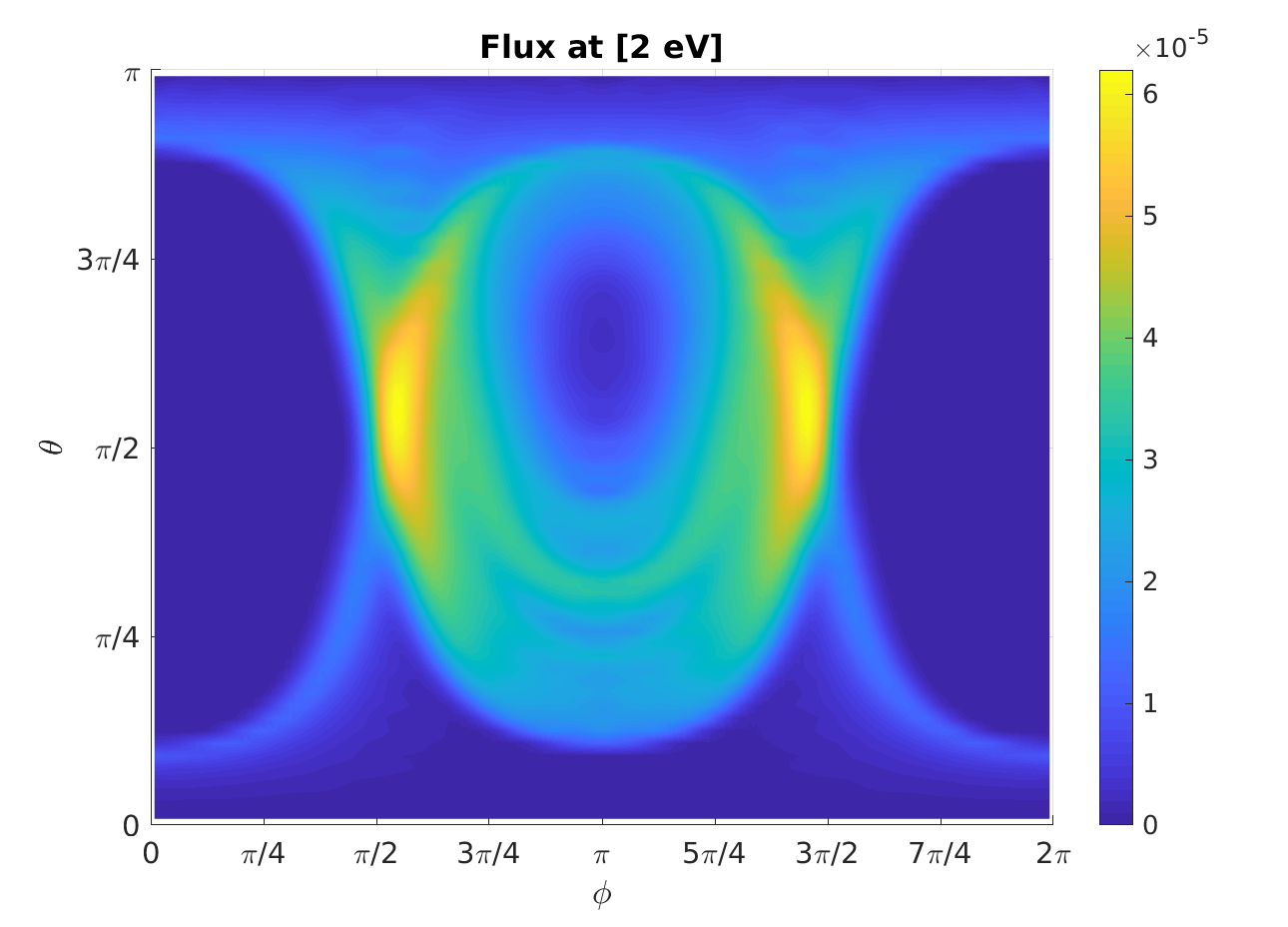}
\includegraphics[width=0.45\linewidth]{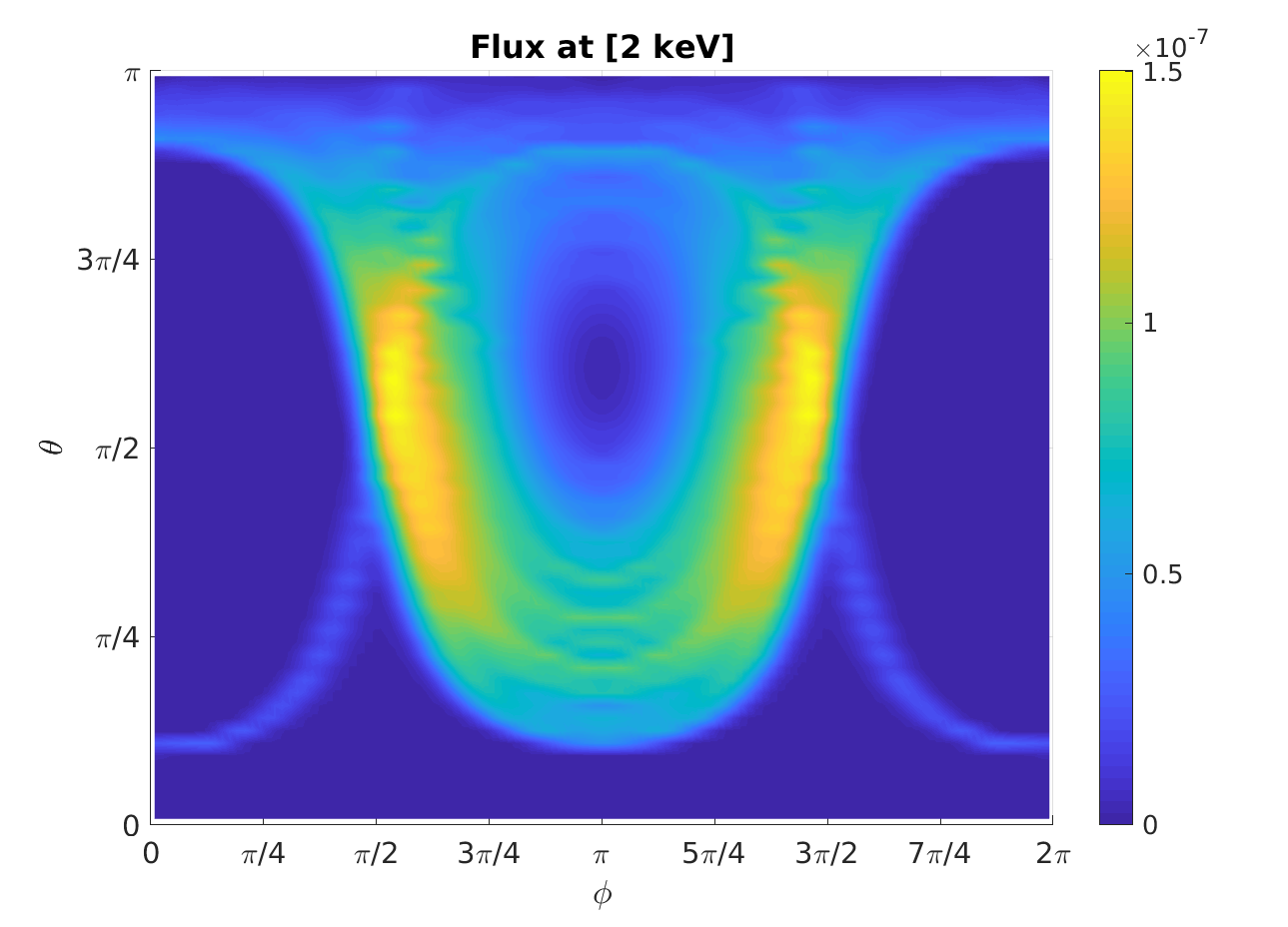}
\includegraphics[width=0.45\linewidth]{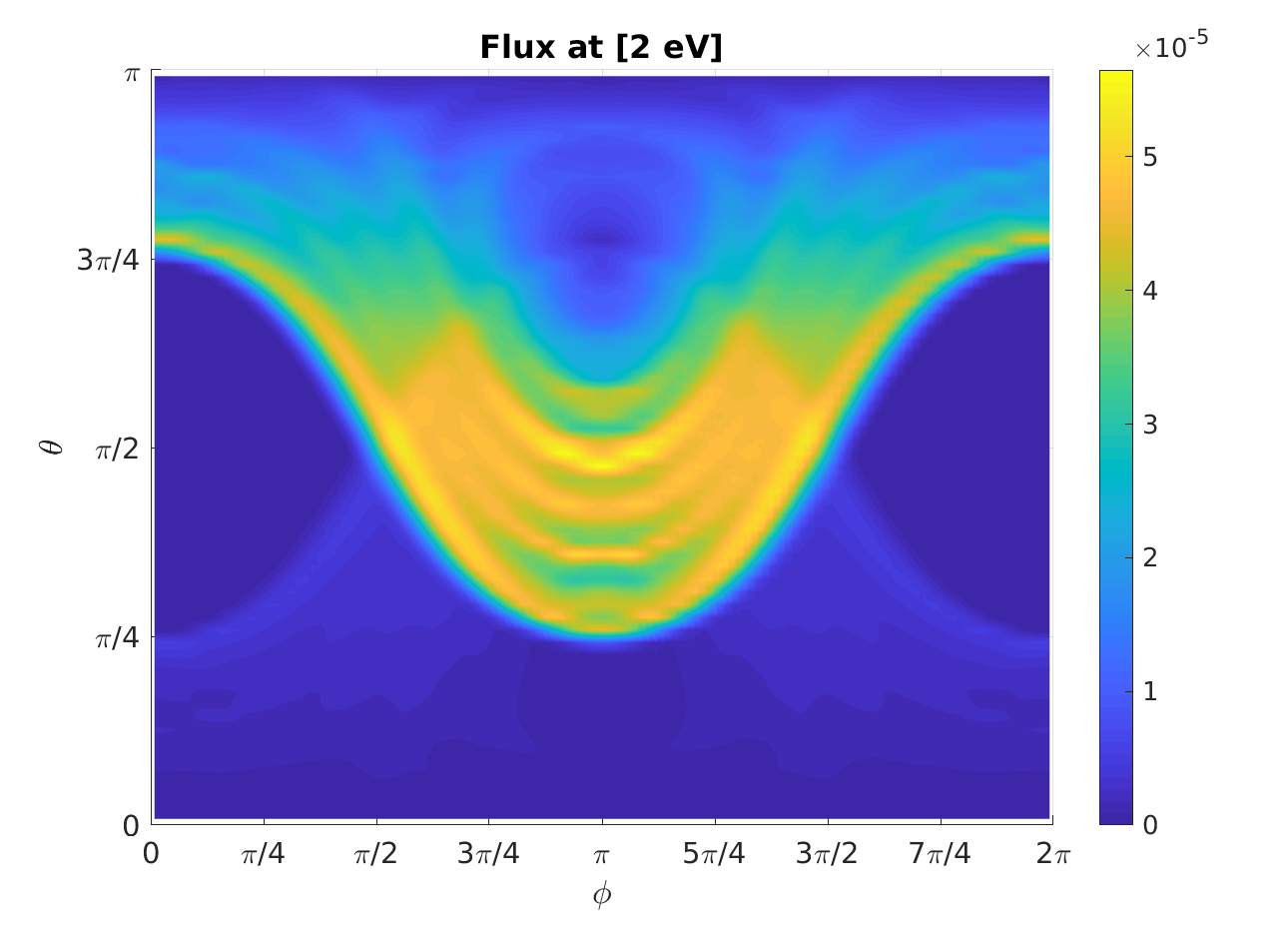}
\includegraphics[width=0.45\linewidth]{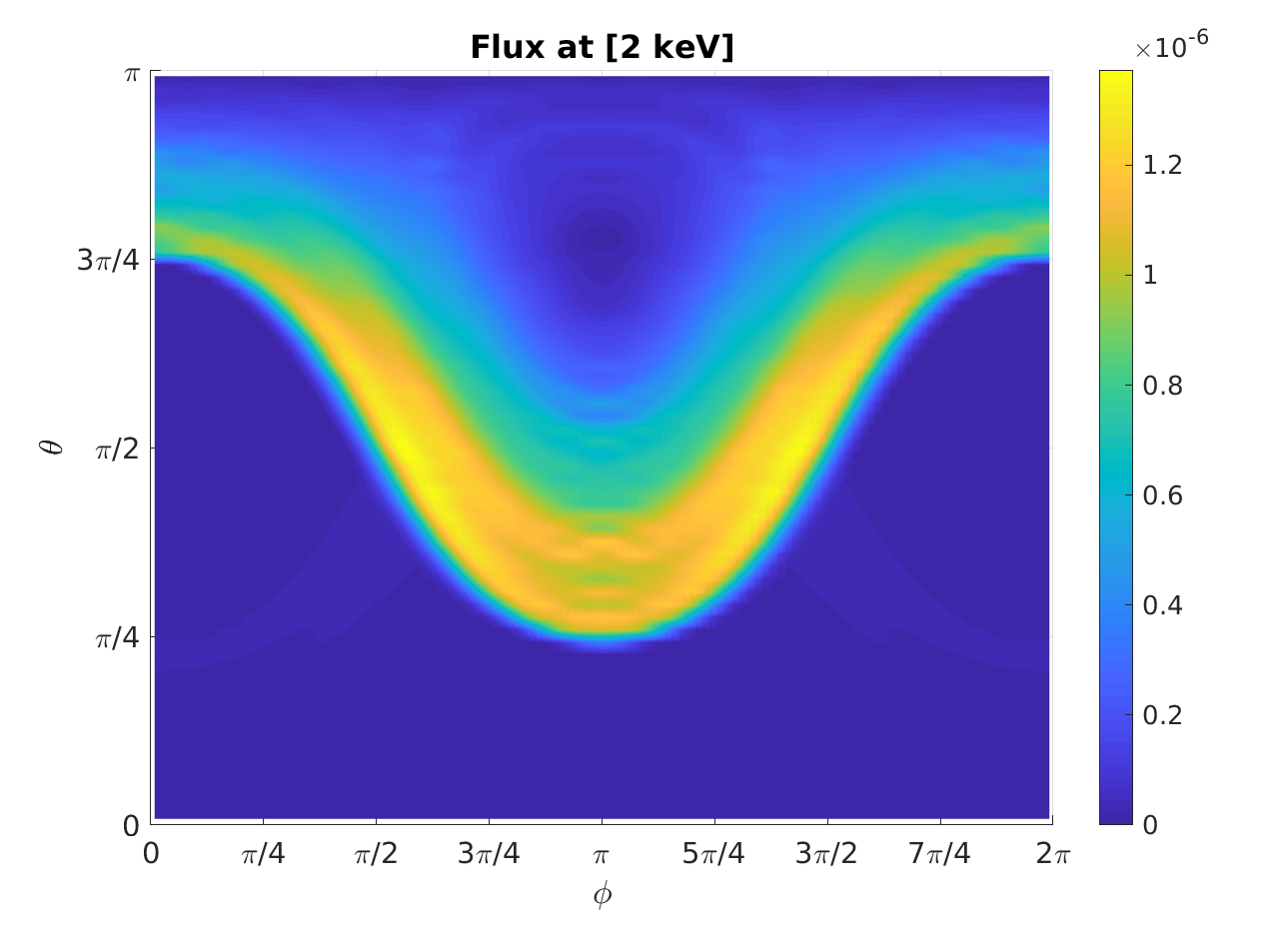}
\includegraphics[width=0.45\linewidth]{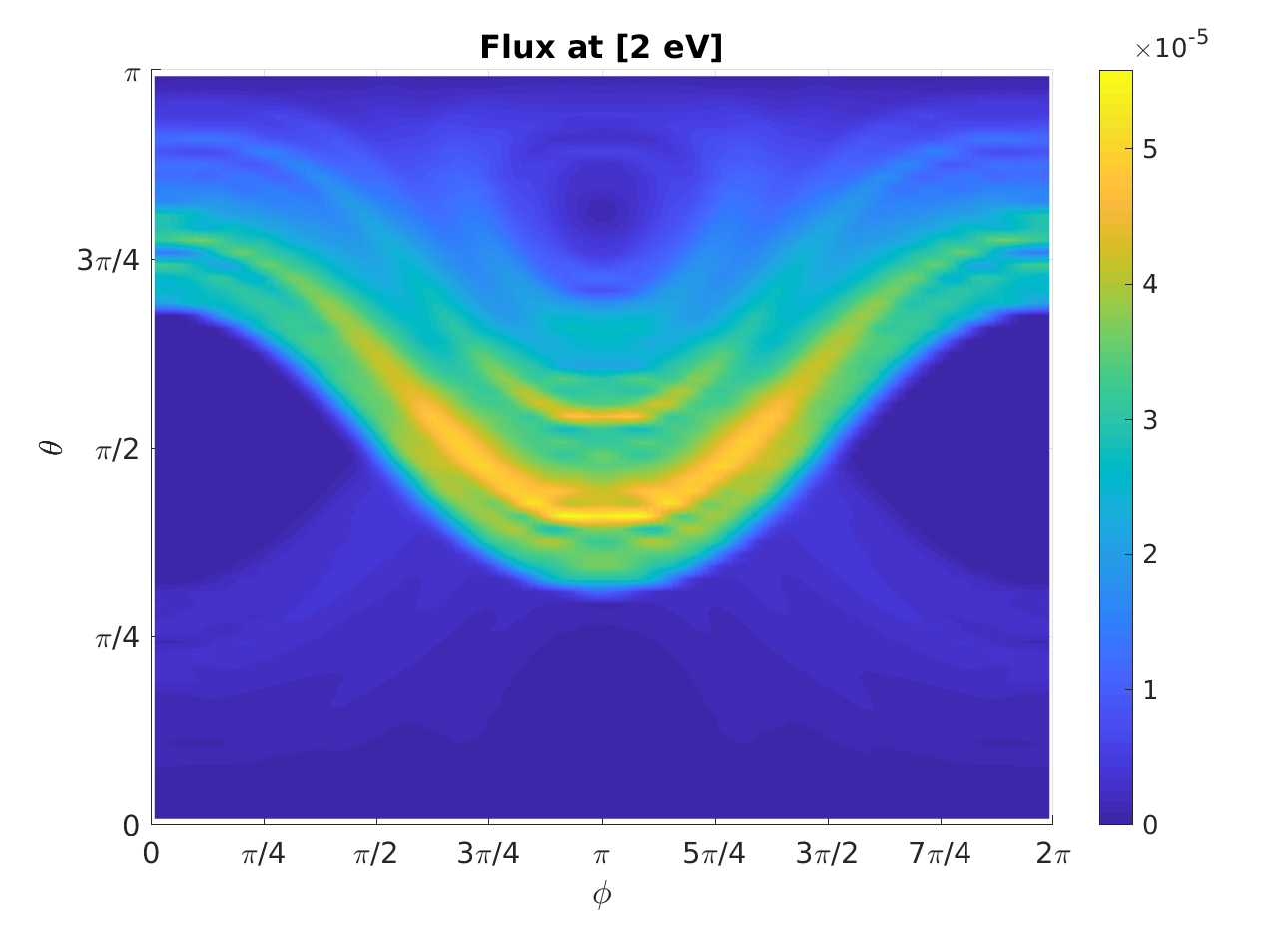}
\includegraphics[width=0.45\linewidth]{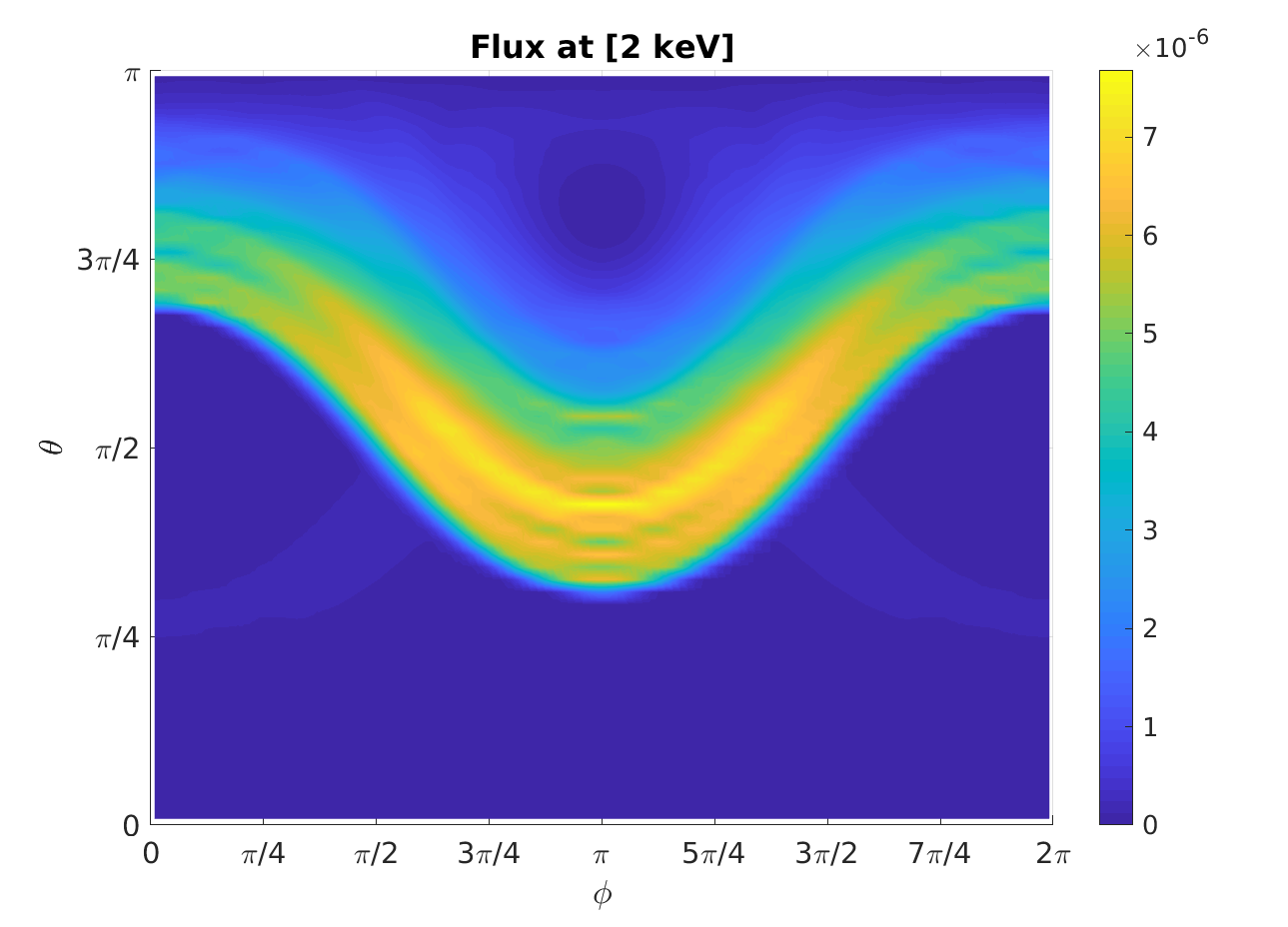}
\caption{Synchrotron Emission maps for various injection locations $\theta_c = \pi/3;\; \pi/4;\; \pi/6;\; \pi/8$ from top to bottom   The synchrotron luminosity maps in optical (left) and in X-ray (right) energy band depends on viewing angle ($\theta,\phi$). Here $\theta = 0$ in direction of magnetic axis, $\phi = 0$  
in direction of magnetic field line plane. For the $\eta_0  = 1.25\times10^{-4}$. }
 \label{fig:lummapsbig}
\end{figure*}

The emission point can be significantly shifted relative to the magnetic equator, so we plot emission patterns for different emission points on the Fig.~\ref{fig:lummapsbig} for $\eta_0 =1.25\times10^{-4}$. As we can see, the light curve can have two peaks or be quite faint. We should note, what the distance between peaks can vary in wide range from $\pi$ up to $0$--complete merge. Also for nonsymmetrical injection point the slopes of the light-curve are different (not symmetric) and if one have fast-rise slow decay the second will have slow-rise and fast-decay, such combination take place if injection point in opposite hemisphere compare to direction to observer. The opposite combination of rise -decay is also possible.

In Fig. \ref{fig:lummapsbig} we investigate various injection locations $\theta_c = \pi/3;\; \pi/4;\; \pi/6;\; \pi/8$. For injection closer to the axis (smaller  $\theta_c $) particles  experience strong in one-go. Their emission pattern is affected by mildly relativistic parallel velocity. 
Qualitatively, this corresponds to trajectory 2   in Fig. \ref{ftrapped-vanAllen1-2}.

\section{Applications and discussion}

Our results have direct implementation to the observed properties of the intermediate polars. 

The direct heating by synchrotron photons un able to form observed compact polar hotspots on the WD \citep{2021ApJ...908..195G}.
On the over hand, the frozen electrons with small initial pitch angle can penetrate deep into \ms and heat the polar hats of the WD as it is discussed in the \citep{2021ApJ...908..195G}.

An interesting point which we get in our calculations is the maximal emission take place at quite large pitch angles (see Fig.~\ref{fig:galum}, $\alpha\sim \pi/4$), so the emission cone is wide for the majority of particles. It can be relatively narrow ($\sim0.1$) only for small initial pith angles, indeed ($\alpha<0.01$). We can say that for small initial pitch angles, the emissivity cone will be at least an order of magnitude wider when initial ones \citep[but see simplified approach in][]{2022MNRAS.510.2998D}.   

We can describe several possible configurations and characteristic pulse profile behavior.
\begin{itemize}
    \item {\it Magnetic moment parallel to the rotation axis and injection is in equatorial plane:} The light curve can have single peak ($\eta_e>10^{-3}$) or close to constant emission, especially in hard energy range.
    \item {\it Magnetic moment parallel to the rotation axis, but injection point varies depends on orbital phase:} depends on observer position light curve can vary from two distinct peaks with smooth deep in between till no signal (weak signal) or narrow single peak. In the case of strong inclination of injection point, the peaks can "merge" to single wide one with steep rise and decay
    \item {\it Magnetic moment not parallel to the rotation axis, and injection point varies depends on orbital phase:} The light curve can have two different in intensity peaks. The peaks should be asymmetric, with relatively shallow fronts between them and steep profiles for opposite fronts.  
\end{itemize}

\acknowledgments

BMV appreciate for partial support BASIS foundation grant \#24-1-2-25-1.

\bibliographystyle{JHEP}

\bibliography{BibTex,reference}    


\end{document}